%% file: milne_SJ_TE.tex
\documentclass[aps,12pt,notitlepage,nofootinbib]{revtex4-1}
\input my

\usepackage{amsmath}
\usepackage{amssymb}

\begin{document}

\title{Perturbative Vacuum Wavefunctional for Gauge Theories
in the Milne Space}
\author{Sangyong Jeon, Thomas Epelbaum}
\affiliation{Department of Physics, McGill University\\
3600 University Street\\
Montr\'eal QC, H3A-2T8\\
Canada}

\begin{abstract}
The spectrum of vacuum fluctuations in the Milne space ({\it i.e.} the $\tau-\eta$ coordinate system)
is an important ingredient in the thermalization
studies in relativistic heavy ion collisions.
In this paper, the Schr\"odinger functional for the gauge theory
perturbative
vacuum is derived for the Milne space.
The Wigner-transform of the corresponding vacuum density functional is also
found together with the propagators.
We finally identify the fluctuation spectrum in vacuum, and show the
equivalence between the present approach and the symplectic product based
method \cite{Dusling:2011rz,Epelbaum:2013waa}.
\end{abstract}

\maketitle

\section{Introduction}

When two heavy ions collide at ultra-relativistic energies,
the collision produces a system that will eventually
evolve into quark-gluon plasma.
In the CGC-Glasma picture of the initial stage
\cite{Iancu:2002xk, Kovner:2005pe, JalilianMarian:2005jf,
Weigert:2005us,Lappi:2006fp,
McLerran:2010ub, Gelis:2010nm, Balitsky:2010jf,
Kovchegov:2012csa, Iancu:2012xa, Gelis:2012ri},
each nucleus is initially composed of two parts; a thin disk made up of
large $x$
partons acting as a static colour source and the strong static gluon field
generated by the disk. Gluon quanta in this gluon field are the small $x$
partons
in the nuclear parton distribution.
When the nuclei collide, the large $x$ partons mostly
pass through each other almost unscathed
(except occasionally producing hard jets)
but the gluon fields from each nuclei start to interact strongly with each
other.

In the lab frame, the two Lorentz contracted nuclei start to overlap at
$t \approx z \approx 0$. At any $t > 0$, the system is then composed of the
two disks located at $z \approx \pm t$ and the gluon field stretched
between them.
In a very simplified picture, one could think of the gluonic system as a
uniform `string'
that is being stretched between two ends pulling away with the speed of
light.
In this `string' picture,
a segment of string at position $z$ at time $t$ will have the speed $v =
z/t$.
Since the length of the string is linearly proportional to time $t$,
this speed will remain constant for any given segment of the string.
This segment therefore can be uniquely labelled by the space-time rapidity
\be
\eta = \tanh^{-1}(z/t)
\ee
and its local time is the proper time
\be
\tau = t/\gamma = \sqrt{t^2 - z^2}
\ee
which are nothing but the Milne coordinates.
Therefore, this coordinate system
is the most natural one to describe the evolution of such systems.
Note that at $\tau = 0$ the spacetime rapidity $\eta$ is not well defined.
Accordingly, theory of the created matter must be
formulated in this space with the restriction that $\tau$
is strictly non-zero and positive definite.

The usual perturbation theory in Minkowski space is concerned about
calculating the scattering amplitude
$\calM = \bra{\rm out}\hatT\ket{\rm in}$
where $\hatT$ is the scattering operator,
the $\ket{\rm in}$ state is defined at the remote past
$t= -\infty$,
and the $\ket{\rm out}$ state is defined at the remote future $t =
+\infty$.
Hence, the problem being solved is not an initial value problem but
a boundary value problem.
In contrast, the quantum field theory in the Milne space is most naturally
formulated as an initial value problem since $\tau$ is restricted to be
positive, and also $\tau = 0^+$ (the trajectory of the colour sources)
represents an actual physical boundary
in any ultra-relativistic heavy ion collisions.

The initial value problem in quantum mechanics is most naturally formulated
in terms of the expectation value
$\ave{\hatO(\tau)} = \bra{\rm in}U^\dagger(\tau)\hatO U(\tau)\ket{\rm in}$
where $\hatU(\tau)$ is the time evolution operator.
Computation of the expectation value
requires the path integral on the
closed time path (CPT)\cite{Schwinger:1960qe,Keldysh:1964ud}.\footnote{%
This formalism is variously known as the in-in formalism, Schwinger-Keldysh
formalism,
and also Keldysh-Schwinger formalism.}
Explicitly,
\be
\ave{\hatO(\tau)}
& = &
\int [d\phi^f]
\int^{\phi^f} \calD\phi_1
\int^{\phi^f} \calD\phi_2\,
\rho_0[\phi_1^0, \phi_2^0]\,
e^{ i\int_{\tau_0}^{\tau_f} d\tau\, \int d^3 \tildex\, \left(\calL(\phi_1)
- \calL(\phi_2) \right) }
\,O[\phi_1,\phi_2]
\label{eq:aveO}
\ee
where
\be
\rho_0[\phi_1^0, \phi_2^0]
=
\braket{\phi_1^{0}}{\rm in}
\braket{\rm in}{\phi_2^{0}}
\ee
is the matrix element of the initial density operator
$\hatrho_0 = \ket{\rm in}\bra{\rm in}$.
The details of $O[\phi_1,\phi_2]$ depends on the time-ordering
structure of the operators in $\hatO$.
The field $\phi_1$ lives on the forward going time line and the field
$\phi_2$ lives on the
backward going time line.
Here $\tau_0 > 0$ is the initial time, and all the quantities with a $0$
subscript are evaluated at $\tau_0$. 
The two fields $\phi_{1,2}$ share the same boundary value $\phi_f$
at the final time which is then traced.
Here, the functional integral measure $\calD\phi$ represents integrating over
a function of both space and time and $[d\phi^f]$ represents integrating over
a function of space at a fixed time (in this case $t = t_f$).

Throughout this paper,
we denote the position and the position integral in the (longitudinal)
Milne space with a tilde.
That is, $\tildex = (\bfxperp, \eta)$ and $d^3\tildex = d^2x_\perp d\eta$.
The subscript $\perp$ denotes only the $x$ and $y$ components
of a vector quantity.

To compute $\ave{\hatO(\tau)}$ in Eq.(\ref{eq:aveO}), it is crucial
to have the initial wave-functional $\braket{\phi_1^{0}}{\rm in}$,
equivalently the matrix element $\rho_0[\phi_1,\phi_2]$.
The goal of this paper is to derive
the initial wave-functional for the perturbative Abelian gauge theory
vacuum in the Milne space. An initial attempt in this direction
was made in \cite{Fukushima:2006ax}. 
Within perturbation theory, non-Abelian gauge
theory vacuum may be obtained by having $N_c^2 - 1$ copies of the Abelian
vacuum. We also verify that the propagators obtained in this way coincide
with those obtained by solving the linearized equation of motion for the
vacuum spectrum of fluctuations, the solutions being normalized with the
help of the natural symplectic product
\cite{Dusling:2011rz,Epelbaum:2013waa}.

In the context of relativistic heavy ion collisions,
the initial state $\ket{\rm in}$ is composed of the classical
Yang-Mills field right off the light-cone and the zero-point
vacuum
fluctuations\cite{Gelis:2006cr,Gelis:2006yv,Gelis:2007pw,Gelis:2008sz}.
In this paper, we concentrate on the vacuum part.
For the classical Yang-Mills part,
see\cite{McLerran:1993ni, McLerran:1993ka, McLerran:1994vd, Ayala:1995kg,
Ayala:1995hx}.
The problem of
vacuum fluctuations in the classical background field is a more difficult
one \cite{Dusling:2011rz,Epelbaum:2013waa}
to which we hope to come back in a future work.

\section{Keldysh rotation and Classical-Statistical Approximation}

The role of the vacuum density functional $\rho_0[\phi_1^{0},\phi_2^{0}]$
can be better elucidated if one makes the Keldysh rotation
\be
&\displaystyle \phi_r = {\phi_1 + \phi_2\over 2}
\\
& \phi_a = \phi_1 - \phi_2
\ee
Equivalently, $\phi_1 = \phi_r + \phi_a/2$ and $\phi_2 = \phi_r -
\phi_a/2$.
The generating functional can be obtained by
expanding $\calL(\phi_1)-\calL(\phi_2)$ in powers of $\phi_a$
\be
\lefteqn{ Z[J_r, J_a] } &&
\non
& = &
\int \calD\phi_r
\int \calD\phi_a\,
\rho_W[\phi_r^0, \pi_r^0]\,
\exp\left(
i\int_{\tau_0}^{\tau_f} d\tau\, d^3\tildex\,
\left(
\phi_a E[\phi_r]
+ {\phi_a^3\over 24} {\partial^3\calL(\phi_r)\over \partial\phi_r^3}
+ J_r\phi_r + J_a\phi_a
\right)
\right)
\non
\label{eq:Zra_in_2}
\ee
where $E[\phi^r] = \int {\delta\calL(\phi_r)/\delta\phi_r}$ is the
classical field
equation, the canonical momentum field is defined as
$\pi_r = \partial\calL/\partial(\partial_\tau\phi_r)$,
We also have performed the functional Wigner transform
\be
\rho_W[\phi_r^0,\pi_r^0]
=
\int [d\phi_a^0]\,
\rho_0\left[\phi^0_r+\frac{\phi^0_a}{2},\phi^0_r-\frac{\phi^0_a}{2}\right]
e^{-i\int d^3\tildex\, \phi_a^0\pi_r^0}
\ee
where the $\phi_a^0\pi_r^0$ term results from the boundary terms
when integrating by part the terms containing $\partial_\tau\phi_a$ in
$\calL$. The functional integration measure $[d\phi_a^0]$ denotes that the integration is performed
with a fixed time, in this case at the initial time $\tau_0$.

The expression Eq.(\ref{eq:Zra_in_2}) looks deceptively simple,
but it needs a careful consideration before it can be useful.
Notice that there is no integration over $\pi_r^0$ while $\rho_W$ depends
on
it. Recall that $\pi_r^0 =
\partial\calL(\phi_r)/\partial(\partial_\tau\phi_r)$.
Hence $\pi_r^0$ contains $\partial_\tau\phi_r$ evaluated at $\tau_0$.
This implies that $\rho_W$ is a functional of $\phi_r^0$ and
$\partial_\tau\phi_r(\tau_0)$.
But how does one interpret the time derivative $\partial_\tau\phi$ in a
path integral?
This problem is not specific to this particular formulation.
In any path integral formulation of
quantum mechanical system, the ``time derivative'' makes sense only in two
ways.
One is to discretize the time so that $\partial_\tau\phi \to
(\phi_{n+1}-\phi_n)/\Delta\tau$.
The second way is to do perturbation theory
using a {\em classical solution} as the zero-th order solution.

When time is discretized
$\rho_W[\phi_r^0, \pi_r^0]$ then provides
restrictions on the first two field values
$\phi_r(\tau_0)$ and $\phi_r(\tau_0+\Delta\tau)$.
This would be the case if one is to evaluate the path integral numerically.
This is a fascinating subject by itself, but in this study we are more
interested in
the perturbative approach without discretizing the time.
If we treat the $\phi_a^3\calL'''/24$ term as a perturbation, then the
unperturbed
part of the generating functional is
\be
Z_0[J_r, J_a]
& = &
\int [d\phi_r^0][d\pi_r^0]\,
\rho_W[\phi_r^0, \pi_r^0]\,
\exp\left(
i\int_{\tau_0}^{\tau_f} d\tau\, d^3\tildex\, J_r
\phi_{\rm cl}[\phi_r^0,\pi_r^0,J_a]
\right)
\label{eq:Z0_in_2}
\ee
using the fact that integration over $\phi_a$ produces
$\delta[E[\phi_r] + J_a]$ that enforces the classical equation of motion.
Hence,
$\phi_{\rm cl}[\phi_r^0,\pi_r^0,J_a]$ here is the {\em classical solution}
with the initial conditions given by $\phi_r^0$ and $\pi_r^0$.
For more details such as converting the boundary value problem
to the initial value problem, see Ref.\cite{Jeon:2004dh}.

If one can find an exact solution to the full nonlinear classical field
equation,
then the evaluation of an expectation value of any operator
$\hatO$ can proceed by taking appropriate
functional derivatives of $Z_0[J_r,J_a]$ to match the expansion of
$\hatO e^{i\int \phi_a^3\calL'''(\phi_r)/24}$ in powers of $\calL'''$.
In this perturbative approach, the time evolution is purely classical.
Quantum effects are contained in the quantum fluctuations in
$\rho_W[\phi_r^0,\pi_r^0]$
and in the quantum correlation in the perturbing Lagrangian
$\phi_a^3\calL'''(\phi_r)/24$.
This is the starting point of the classical-statistical approximation.

If one cannot find an exact solution, then one needs to expand the
interaction term
in $\phi_a E[\phi_r]$ as well. In that case, $Z_0[J_r,J_a]$ becomes the
generating
functional for the free-field and the resulting perturbation theory is the
usual
one with the free-field propagators.
It is worth noting that the time evolution in a free-field theory is
completely
classical.
Quantum effects reside solely in the fluctuations of the initial values.

{}From Eq.(\ref{eq:Z0_in_2}), it is clear that the Wigner transform
$\rho_W[\phi_r^0,\pi_r^0]$ can play the role of the probability density
functional for the initial field values.  Of course, as a Wigner
functional,
it is {\em not} guaranteed that $\rho_W$ is always positive definite.
Fortunately as we show in this paper, the perturbative vacuum state
usually results in a positive definite $\rho_W$ simply because it is
usually
a gaussian.
It is also clear from Eq.(\ref{eq:Z0_in_2}) that
the two most important ingredients in obtaining $Z_0[J_r, J_a]$
are the Wigner functional $\rho_W[\phi_r^0,\pi_r^0]$
and the full classical solution
$\phi_{\rm cl}[\phi_r^0,\pi_r^0,J_a]$.

\section{Summary of Results}

Since this paper is of necessity technical, it will be useful to list the
main results
here before going into the details of calculations.
Our gauge choice is $A_\tau = 0$. The spatial gauge vector field
is further decomposed into
the transverse vector $\bfA_\perp$ and the longitudinal component $A_\eta$.
The 2-d vector $\bfA_\perp$ is further decomposed into
$A_T(\tildek) = \hatk^x A^y - \hatk^y A^x$ with $\hatbfk_\perp =
\bfk_\perp/k_\perp$,
and $\varphi$ defined by $\nabla_\perp{\cdot}\bfA_\perp =
\nabla_\perp^2\varphi$.

The transverse component $A_T$ satisfies the Klein-Gordon equation in the
Milne space.
Hence, we can adapt the results for the scalar theory found
in Refs.\cite{Long:1996wf,Long:1996bk}.
The vacuum functional for $A_T$, re-derived in Section
\ref{sec:TransverseVac}, is known to
be (omitting the $1,2$ indices of the Schwinger-Keldysh contour)
\be
\braket{A_T}{\rm vac}
=
\calN_T\,
\exp\left(
-{1\over 2}\int {d^3\tildek\over (2\pi)^3}\,
A_T(-\tildek)
\left(
 {-i\Pi^{(1)}_{ik_\eta}(k_\perp \tau_0)
\over H_{ik_\eta}^{(1)}(k_\perp\tau_0)}
\right)
A_T(\tildek)
\right)
\ee
Here $\calN_T$ is the normalization factor and $H^{(1)}_{ik_\eta}(x)$ is
the Hankel
function of the first kind and of imaginary order, and we defined
$\Pi_{ik_\eta}^{(1,2)}(x) = x\partial_x H^{(1,2)}_{ik_\eta}(x)$.
The Wigner transform of the density operator $\hatrho_0 = \ket{\rm
vac}\bra{\rm vac}$
is in the retarded-advanced basis
\be
\rho_W^T[A_{rT}^0, \pi_{rT}^0]
=
\calN_T'
\exp\left(
-\int {d^3\tildek\over (2\pi)^3}\left[ R_T(\tau_0,\tildek)
|A_{rT}^0(\tildek)|^2
+
{|\pi_{rT}^0(\tildek) + I_T(\tau_0,\tildek)A_{rT}^0(\tildek)|^2
\over R_T(\tau_0,\tildek)}\right]
\right)
\non
\label{eq:rho_WT_0}
\ee
where $R_T$ and $I_T$ are the real and the imaginary parts
of ${-i\Pi^{(1)}_{ik_\eta}(k_\perp \tau_0)
/ H_{ik_\eta}^{(1)}(k_\perp\tau_0)}$, respectively.
Eq.(\ref{eq:rho_WT_0}) is a new result.
The Wigner function $\rho_W^T[A_{rT}^0,\pi_{rT}^0]$ represents the
zero-point motion of the
quantum fields in vacuum whose effect manifests as the
fluctuations of $A_{rT}^0$ and $\pi_{rT}^0$ which are the initial values of
the $A_{rT}$ field
evolution.
It also plays the role of the probability density functional
of $A_{rT}^0$ and $\pi_{rT}^0$ in the context of the classical-statistical
approximation.
In numerical implementations, $\tau_0$ needs to be kept finite since
neither
$R_T(\tau_0,\tildek)$ nor $I_T(\tau_0,\tildek)$ have a well defined limit
as $\tau_0\to 0$.

The $r$-$a$ propagator is the retarded Green function of the d'Alembertian
in the Milne space. It does not depend on $\rho_W^T$.
It is found in Section \ref{sec:transverse_props} to be
\be
\ave{A_{rT}(\tau,-\tildek)A_{aT}(\tau',\tildek)}
&= &
iG_R(\tau,\tildek)
\non
& = &
{\pi \over 4}
\theta(\tau-\tau')
\left(
H_{ik_\eta}^{(1)}(k_\perp\tau)
H_{ik_\eta}^{(2)}(k_\perp\tau')
-
H_{ik_\eta}^{(2)}(k_\perp\tau)
H_{ik_\eta}^{(1)}(k_\perp\tau')
\right)
\ee
The symmetric propagator is found to be
\be
\ave{A_{rT}(\tau,-\tildek)A_{rT}(\tau',\tildek)}
& = &
{\pi\over 8}
\left(
H_{ik_\eta}^{(1)}(k_\perp\tau)
H_{ik_\eta}^{(2)}(k_\perp\tau')
+
H_{ik_\eta}^{(2)}(k_\perp\tau)
H_{ik_\eta}^{(1)}(k_\perp\tau')
\right)
\ee
These propagators coincide with the Minkowski space retarded and the
symmetric propagators
as shown in Appendix \ref{app:Mprops}.
The exact same forms apply to the single component scalar field theory.

For the longitudinal part, the two components $A_\eta$ and $\varphi$
couples.
The fields $A_\eta$ and $\varphi$ themselves do not satisfy the free-field
equation,
but their conjugate momentum fields
$\pi_\eta = (1/\tau)\partial_\tau A_\eta$ and
$\pi_\varphi = -\tau\nabla_\perp^2\partial_\tau \varphi$ do.
Using this and the fact that the Gauss law requires
$\partial_\eta\pi_\eta = \pi_\varphi$,
it is shown in Section \ref{sec:LongiVac} that the vacuum functional is
given by (omitting the $1,2$ indices of the Schwinger-Keldysh contour)
\be
\braket{A_\eta,\varphi}{\rm vac}
&= &
\calN_L\,
\exp\left(
-{1\over 2}\int {d^3\tildek\over (2\pi)^3}\,
\alpha_\eta(-\tildek)
\left( {
k_\perp^2 H_{ik_\eta}^{(1)}(k_\perp\tau_0)
\over
-i\Pi^{(1)}_{ik_\eta}(k_\perp \tau_0) }
\right)
\alpha_\eta(\tildek)
\right)
\ee
where $\alpha_\eta(\tildek) = A_\eta(\tildek) - ik_\eta\varphi(\tildek)$.
The Wigner transform of the vacuum density functional is found to be in the
retarded-advanced basis
\be
\rho_W^L[A_{r\eta}^0, \pi_{r\eta}^0]
=
\calN_L'
\exp\left(
-\int {d^3\tildek\over (2\pi)^3} \left[R_\eta(\tau_0,\tildek)
|A_{r\eta}^0(\tildek)|^2
+
{|\pi_{r\eta}^0(\tildek) + I_\eta(\tau_0,\tildek)A_{r\eta}^0(\tildek)|^2
\over R_\eta(\tau_0,\tildek)}\right]
\right)
\non
\ee
where the real and the imaginary parts of
${
k_\perp^2 H_{ik_\eta}^{(1)}(k_\perp\tau_0)
/ (-i\Pi^{(1)}_{ik_\eta}(k_\perp \tau_0))
}$ have been denoted $R_\eta(\tau_0,\tildek)$ and $I_\eta(\tau_0,\tildek)$,
respectively.
Again, this can be interpreted as the probability density of initial values
$A_{r\eta}^0(\tildek)$ and $\pi_{r\eta}^0(\tildek)$
which are the initial values for the $A_{r\eta}$ field evolution.
For the initial values of the $\varphi_r$ field,
$\varphi_{r}^0(\tildek)$ can be
set to zero using the residual gauge freedom and
$\pi_{r\varphi}^0(\tildek)$ is given by $ik_\eta\pi_{r\eta}^0(\tildek)$.
Note that having
$\varphi_r(\tau_0,\tildek) =\varphi^0_r(\tildek)= 0$ does not imply
$\varphi_1$ and $\varphi_2$
both vanish at $\tau_0$. Only their sum is required to vanish.

The longitudinal propagators in the $r{-}a$ formalism
are found in Section \ref{sec:longi_props} to be
\be
\ave{A_{r\eta}(\tau,\tildek)A_{a\eta}(\tau',\tildek)}
& = &
-i
\theta(\tau-\tau_0)
\int_{\tau_0}^{\tau}d\tau''\tau''\,
\left(\tau' \partial_{\tau'}G_R(\tau'',\tau';\tildek)
+ \tau_0 G_R(\tau'',\tau_0;\tildek)\delta(\tau'-\tau_0)
\right)
\non
\ee
and the symmetric propagator
\be
\lefteqn{
\ave{A_{r\eta}(\tau,\tildek)A_{r\eta}(\tau',-\tildek)}
} &&
\non
& = &
{\pi \Pi^{(1)}_{ik_\eta}(k_\perp\tau_0) \Pi^{(2)}_{ik_\eta}(k_\perp\tau_0)
\over 4 k_\perp^2}
+ {\pi k_\perp^2\over 8}
\left(
R^{(1)}(\tau,\tildek)R^{(2)}(\tau',-\tildek)
+
R^{(2)}(\tau,\tildek)R^{(1)}(\tau',-\tildek)
\right)
\non && {}
-{\pi\over 8}
\left(
\Pi^{(2)}_{ik_\eta}(\tau_0)\left[R^{(1)}(\tau,\tildek)
+
R^{(1)}(\tau',-\tildek)\right]
+
\Pi^{(1)}_{ik_\eta}(\tau_0)\left[R^{(2)}(\tau,\tildek)
+
R^{(2)}(\tau',-\tildek)\right]
\right)
\ee
with
\be
R^{(1,2)}(\tau,\tildek)
& = &
\int_{\tau_0}^{\tau}d\tau' \tau'\, H^{(1,2)}_{ik_\eta}(k_\perp\tau')
\ee
These are all new results. Now we turn to the calculation of these results.

\section{Abelian Gauge Theory in the Milne Space}

In a curved space, the fundamental $U(1)$ gauge field is a covariant vector $A_a$
and the field strength is given by
\be
F_{ab} = \partial_a A_b - \partial_b A_a
\ee
The Milne space metric is given by
\be
ds^2 = g_{ab}dx^a dx^b
= -d\tau^2 + \tau^2d\eta^2 + dx^2 + dy^2
\ee
where $t \ge 0$, $\tau^2 = t^2 - z^2 \ge 0$
while $\eta = \tanh(t/z)$ can take any real value.
The Abelian gauge theory Lagrangian in the Milne space is
\be
L & = &
\int d^3\tildex\, \calL
\non
&=& -{1\over 4}\int d^3\tildex\, \sqrt{-g} g^{ac}g^{bd} F_{ab} F_{cd}
\non
& = &
{1\over 2}
\int d^3\tildex\,
\tau\left(
F_{\tau x}^2 + F_{\tau y}^2 + {1\over \tau^2}F_{\tau\eta}^2
- F_{xy}^2 - {1\over \tau^2}F_{x\eta} - {1\over \tau^2} F_{y\eta}^2
\right)
\label{eq:Ltaueta}
\ee
where $g = {\rm det}g_{ab} = -\tau^2$ and we set the coupling constant to
unity.

In the CPT formalism, the generating functional
of the propagators is composed of
the forward going time line and the backward going time line
\be
\lefteqn{Z_{\rm free}[J_1, J_2]}&&
\non
& = &
\int [dA^f][dA^{0}_1][dA^{0}_2]\,
\rho_{\rm vac}[\tau_0, A^{0}_1, A^{0}_2]\,
\int_{A^{0}_1}^{A^f}\calD A_1
\int_{A^{0}_2}^{A^f}\calD A_2
e^{
i\int_{\tau_0}^\infty d\tau\int d^3\tildex
\left(\calL_{\rm free}(A_1) - \calL_{\rm free}(A_2)
+ J_1 A_1 - J_2 A_2 \right)
}
\non
\label{eq:Zfree}
\ee
Here $A_1$ is the field on the forward going time line and
$A_2$ is the field on the backward going time line.
For obtaining the propagators only the quadratic part of the Lagrangian
$\calL_{\rm free}(A)$ is needed.
The fields start with the initial values $A_1^0$
and $A_2^0$
at the initial time $\tau_0$ and share the common final value $A^f$ which
is then traced.
Note that
the Lorentz indices and space-time dependence
are suppressed in this expression for simplicity of notation.

In terms of the ground-state wave-functional
$\Psi_0[\tau, A^{0}_1] = \braket{A^{0}_1}{\rm vac, \tau}$,
the vacuum functional is given by
\be
\rho_{\rm vac}[\tau, A^{0}_1, A^{0}_2]
& = &
\braket{A^{0}_1}{\rm vac, \tau}\braket{\rm vac, \tau}{A^{0}_2}
=
\Psi_0[\tau, A^{0}_1]\Psi_0^*[\tau,A^{0}_2]
\ee
Our task here is then to obtain $\Psi_0[\tau, A^{0}_1]$
that satisfies the functional Schr\"odinger equation
\cite{Greensite:1979yn,Jevicki:1979mb,Jackiw:1980xj}
\be
i\partial_\tau \Psi_0[\tau,A] =
\hatH\left[A, \hat\pi\right] \Psi_0[\tau,A]
\ee
where
$\hatH[A, \hat\pi]$ is the functional Hamiltonian with
the functional momentum operator $\hat\pi = -i\delta/\delta A$.
Since the Hamiltonian turns out to be $\tau$ dependent as shown below,
this cannot be reduced to the time-independent Schr\"odinger equation.

To obtain the functional Hamiltonian for an Abelian gauge theory,
we first start with the Lagrangian in Eq.(\ref{eq:Ltaueta}).
In the $A_\tau = 0$ gauge,
the Lagrangian density in Eq.(\ref{eq:Ltaueta}) is
(dropping the branch indices 1 and 2 for now)
\be
\calL
& = &
{\tau\over 2} (\partial_\tau \bfAperp)^2
-
{\tau\over 2}
\left(
\nabla_\perp{\times}\bfAperp
\right)^2
+
{1\over 2\tau}
\bfAperp{\cdot} \partial_\eta^2 \bfAperp
\non
& & {}
+
{1\over 2\tau}
(\partial_\tau A_\eta)^2
+
{1\over 2\tau}
A_\eta \nabla_\perp^2 A_\eta
-
{1\over \tau}
A_\eta \partial_\eta \nabla_\perp{\cdot}\bfAperp
\ee
with the Gauss law constraint
\be
0 &= &
\frac{\partial_\eta\partial_\tau A_\eta}{\tau^2}
+ \nabla_\perp{\cdot}\partial_\tau\bfA_\perp
\label{eq:Gauss0}
\ee
We further decompose the transverse field as follows
\be
\bfA_\perp = \bfA_T + \bfA_L
\ee
where
\be
\bfA_T(\tau, \tildex)
&=&
i\int {d^2 k_\perp\over (2\pi)^2}
(\hatbfk_\perp \times \bfe_z )
A_T(\tau, \eta, \bfkperp) \,
e^{i\bfk_\perp{\cdot}\bfx_\perp}
\\
\bfA_L(\tau, \tildex)
&=&
\nabla_\perp \varphi(\tau, \tildex)
\ee
These satisfy
\be
&\nabla_\perp{\cdot}\bfA_T = 0
\\
&\nabla_\perp{\times}\bfA_L = 0
\\
\inline{and}
&\int d^2 x_\perp\,
\bfA_T(\tau,\tildex){\cdot}\bfA_L(\tau,\tildex)
= 0
\ee
In terms of $A_T$ and $\varphi$,
the Lagrangian density becomes
\be
\calL
& = &
{\tau\over 2} (\partial_\tau A_T)^2
+
{\tau\over 2}
A_T \nabla_\perp^2 A_T
+
{1\over 2\tau} A_T \partial_\eta^2 A_T
\non
& & {}
+
{1\over 2\tau}
(\partial_\tau A_\eta)^2
+
{\tau\over 2} (\partial_\tau \nabla_\perp \varphi)^2
+
{1\over 2\tau}
\left(
A_\eta \nabla_\perp^2 A_\eta
+ \nabla_\perp \varphi {\cdot} \partial_\eta^2 \nabla_\perp \varphi
- 2 A_\eta \partial_\eta \nabla_\perp^2\varphi
\right)
\ee
The momentum fields are given by
\be
{\partial \calL\over \partial (\partial_\tau A_T)}
&= &
\pi_T = \tau \partial_\tau A_T
\\
{\partial \calL\over \partial (\partial_\tau A_\eta)}
&=&
\pi_\eta = {1\over \tau}\partial_\tau A_\eta
\\
{\partial \calL\over \partial (\partial_\tau \varphi)}
&=&
\pi_\varphi = -\tau \nabla_\perp^2 \partial_\tau \varphi
\ee
and the Hamiltonian density is
\be
\calH
& = &
\pi_\eta \partial_\tau A_\eta
+
\pi_\varphi \partial_\tau \varphi
+
\pi_T \partial_\tau A_T
-
\calL
\non
& = &
{1\over 2\tau} \pi_T \pi_T
-
{\tau\over 2}
A_T \nabla_\perp^2 A_T
-
{1\over 2\tau}
A_T \partial_\eta^2 A_T
\non
& & {}
+ {\tau\over 2} \pi_\eta^2
- {1\over 2\tau} \pi_\varphi {1\over \nabla_\perp^2} \pi_\varphi
+
{1\over 2\tau}
\left(
- A_\eta \nabla_\perp^2 A_\eta
+ \varphi \partial_\eta^2 \nabla_\perp^2\varphi
+ 2 A_\eta \partial_\eta \nabla_\perp^2\varphi
\right)
\ee
The Gauss law constraint Eq.(\ref{eq:Gauss0}) becomes
\be
\partial_\eta \pi_\eta = \pi_\varphi
\label{eq:Gauss3}
\ee
{}From the canonical quantization rule
\be
[A_a(\tau,\tildex), \hat\pi_b(\tau,\tildey)] =
i\delta_{ab}\delta(\tildex-\tildey)
\ee
we can identify the momentum-field operators
\be
\hat\pi_T = -i{\delta\over\delta A_T},\
\hat\pi_\eta = -i{\delta\over\delta A_\eta},\
\hat\pi_\varphi = -i{\delta\over\delta \varphi}
\ee
The functional Schr\"odinger equation then takes the following form
\be
\lefteqn{
i\partial_\tau \Psi_0[\tau, A_T, A_\eta, \varphi]
} &&
\non
& = & \hatH\Psi[\tau, A_T, A_\eta, \varphi]
\non
& = &
\int d^3\tildex\,
\Bigg(
-{1\over 2\tau} {\delta^2 \over \delta A_T^2}
-
{1\over 2\tau}
A_T \partial_\eta^2 A_T
-
{\tau\over 2}
A_T \nabla_\perp^2 A_T
\non
& & {}
-{\tau\over 2} {\delta^2 \over \delta A_\eta^2}
+
{1\over 2\tau} {\delta\over \delta\varphi}
{1\over \nabla_\perp^2} {\delta\over \delta\varphi}
+
{1\over 2\tau}
\left(
- A_\eta \nabla_\perp^2 A_\eta
+ \varphi\partial_\eta^2 \nabla_\perp^2\varphi
+ 2 A_\eta \partial_\eta \nabla_\perp^2\varphi
\right)
\Bigg)
\Psi_0[\tau, A_T, A_\eta, \varphi]
\non
\label{eq:Psi0_eq}
\ee
with the Gauss law constraint
\be
\partial_{\eta} {\delta\over \delta A_\eta}
\Psi_0
=
{\delta \over \delta\varphi}\Psi_0
\label{eq:GaussPsi}
\ee
Note that since we are in the Schr\"odinger picture, the functional
coordinates $A_T, A_\eta, \varphi$ does not depend on $\tau$.
What we need in the generating functional $Z_{\rm free}[J_1,J_2]$
is the solution of the Schr\"odinger equation evaluated at
the initial time $\tau = \tau_0$.

\section{Transverse Part of the Vacuum Functional}
\label{sec:TransverseVac}

Since the transverse and the longitudinal
parts separate in Eq.(\ref{eq:Psi0_eq}), the wavefunctional factorizes
\be
\Psi_0[\tau, A_T, A_\eta, \varphi]
=
\Psi_T[\tau, A_T] \Psi_L[\tau, A_\eta, \varphi]
\ee
We consider the transverse part first in this section.
The longitudinal part will be discussed in the next section.

The time dependent Schr\"odinger equation Eq.(\ref{eq:GaussPsi})
is valid for any quantum state, not just the vacuum state.
If the Hamiltonian is time independent (for instance
in the Minkowski time $t$), then we can let
\be
\Psi_T[t, A] = e^{-iEt}\Psi_T[A]
\ee
and solve for the time independent equation
\be
E\Psi_T[A] = H_{\rm Mink.}[A,\hat\pi]\Psi_T[A]
\ee
However in the Milne space, evolution in $\tau$ does not preserve
total energy since the Hamiltonian in Eq.(\ref{eq:Psi0_eq}) is explicitly
$\tau$-dependent. Therefore, the usual concept of vacuum as the lowest
energy eigenstate of the Hamiltonian does not apply.

On the other hand, the Milne space is just a re-parametrization of
the flat Minkowski space.
Therefore, the state representing vacuum should have a similar structure
as the Minkowski vacuum functional.
In the Minkowski space, the perturbative vacuum for a scalar mode is known
to be
\be
\Psi_{\rm Mink.}[t,A]
& = &
\exp\left(-iE_0t -{1\over 2}\int {d^3k\over (2\pi)^3} E_k\,
A(k)A(-k)\right)
\ee
where $E_0 = V\int E_k d^3k/(2\pi)^3$ is the un-renormalized vacuum
energy.
The corresponding state in the Milne space should also have a Gaussian
structure. Furthermore, for the two space-time points inside the Milne
space, they
should result in the same two point correlation functions (propagators) as
in the Minkowski
space.
Hence, we employ the following Gaussian ansatz for the vacuum functional
\be
\Psi_T[\tau,A_T]
& = &
\calN_T(\tau) \exp
\Bigg(
-{\tau^2\over 2} \int d^3\tildex \int d^3 \tildey\,
A_T(\tildex) G_T(\tau,\tildex-\tildey) A_T(\tildey)
\Bigg)
\ee
Since we are in the Schr\"odinger picture,
$A_T(\tildex)$ does not depend on $\tau$ but the kernel
$G_T(\tau,\tildex-\tildey)$ does depend on $\tau$.
Consistency under $\tildex\leftrightarrow \tildey$ exchange
requires
\be
G_T(\tau, \tildex-\tildey)
& = & G_T(\tau, \tildey-\tildex)
\ee
By comparing the left hand side and the right hand side of the functional
Schr\"odinger equation (\ref{eq:Psi0_eq}),
one obtains the equation for the normalization factor
\be
i\partial_\tau\calN_T(\tau) =
{\tau\over 2} \int d^3\tildex\, G_T(\tau, 0)\, \calN_T(\tau)
\ee
and
the equation for the transverse kernel
\be
i\partial_\tau (\tau^2 G_T(\tau,\tildex-\tildey))
={\tau^3}
\int d^3\tildeu\,G_T(\tau, \tildex-\tildeu)G_T(\tau,\tildeu-\tildey)
+\left({\tau}\nabla_\perp^2 + {1\over \tau}\partial_\eta^2\right)\,
\delta(\tildex-\tildey)
\label{eq:GTeq2}
\ee
The normalization factor $\calN_T(\tau)$ obviously involves infinities
since the integrand does not depend on $\tildex$. It requires
renormalization.
However since $\calN_T(\tau)$ is independent of the field variables,
it does not concern us in this paper.

In the momentum space, the equation for $G_T$ simplifies
\be
i\partial_\tau (\tau^2 G_T(\tau,\tildek))
={\tau^3}
G_T(\tau, \tildek)G_T(\tau,\tildek)
-\left({\tau}k_\perp^2 + {k_\eta^2\over \tau}\right)\,
\label{eq:GTeq3}
\ee
Following the Refs.\cite{Long:1996wf,Long:1996bk}, define
\be
-i\partial_\tau \psi_T(\tau,\tildek)
& = &
\tau G_T(\tau,\tildek)\psi_T(\tau,\tildek)
\label{eq:psikdef}
\ee
Multiplying $\psi_T(\tau,\tildek)$ to Eq.(\ref{eq:GTeq3})
and using Eq.(\ref{eq:psikdef}), Eq.(\ref{eq:GTeq3}) becomes
\be
\partial^2 \psi_T = 0
\label{eq:dalembertian}
\ee
where
\be
\partial^2
=
\partial_\tau^2
+
{1\over \tau}
\partial_\tau
+ k_\perp^2 + {k_\eta^2\over \tau^2}
\ee
is the d'Alembertian in the Milne space.

The solutions of Eq.(\ref{eq:dalembertian}) are
the Hankel functions of imaginary order
$H^{(1)}_{ik_\eta}(k_\perp \tau)$
and
$H^{(2)}_{ik_\eta}(k_\perp \tau)$.
If vacuum is to be stable,
the real part of $G_T$ must be positive definite.
Furthermore, it must be able reproduce the Minkowsi correlator
when the two points are both inside the forward light cone.
For these to hold, we must choose
$\psi_T(\tau,\tildek) = c H^{(1)}_{ik_\eta}(k_\perp\tau)$
where $c$ is a constant.
Using Eq.(\ref{eq:psikdef}) one obtains
\be
\tau^2 G_T(\tau,\tildek) = {-i\tau \partial_\tau
H^{(1)}_{ik_\eta}(k_\perp\tau)
\over H^{(1)}_{ik_\eta}(k_\perp\tau)}
\label{eq:tau2GT}
\ee
The real part of $\tau^2 G_T(\tau,\tildek)$ is indeed positive definite.
\be
R_T(\tau, \tildek)
& = &
{\rm Re}(\tau^2 G_T(\tau,\tildek))
\non
& = &
{2\over\pi\, H^{(1)}_{ik_\eta}(k_\perp\tau) H^{(2)}_{ik_\eta}(k_\perp\tau)
} > 0
\ee
where we used the Wronskian of the Bessel functions
\be
H_{ik}^{(1)}(\tau) \partial_\tau H_{ik}^{(2)}(\tau)
- H_{ik}^{(2)}(\tau) \partial_\tau H_{ik}^{(1)}(\tau)
=
-{4i\over \pi \tau}
\label{eq:HWronskian}
\ee
and the fact that
$H^{(2)}_{ik_\eta}(k_\perp\tau) = e^{\pi k_\eta}
(H^{(1)}_{ik_\eta}(k_\perp\tau))^*$
(see Appendix \ref{app:Hankel} for Hankel function properties).
Had we chosen $H^{(2)}_{ik_\eta}(k_\perp\tau)$ as our solution,
the real part of $G_T(\tau,\tildek)$ would have been negative definite.
It is convenient to define
\be
\Pi^{(1,2)}_{ik_\eta}(k_\perp\tau)
=
\tau\partial_\tau H^{(1,2)}_{ik_\eta}(k_\perp\tau)
\ee
as they occur frequently in our analysis.
The fact that $\Psi_T[\tau,A_T]$ is compatible with
$\Psi_{\rm Mink.}[t,A_T]$
is demonstrated in Section \ref{sec:transverse_props} by
showing that the resulting propagators are the same.

The matrix element of the vacuum density operator at $\tau_0$
is now given by
\be
\lefteqn{
\rho_{\rm vac}[\tau_0, A^0_{1T}, A^0_{2T}]
} &&
\non
& = &
\Psi_T[\tau_0, A^0_{1T}]
\Psi_T^*[\tau_0, A^0_{2T}]
\non 
& = &
\exp\left(
-{\tau_0^2\over 2}
\int {d^3\tildek \over (2\pi)^3}\,
\left[A^0_{1T}(-\tildek)G_T(\tau_0,\tildek)A^0_{1T}(\tildek)
+
A^0_{2T}(-\tildek)G_T^*(\tau_0,\tildek)
A^0_{2T}(\tildek)\right]
\right)
\non
\label{eq:rhoT}
\ee

\section{Longitudinal Part of the Vacuum Functional}
\label{sec:LongiVac}

For the longitudinal part, our ansatz is
\be
\Psi_L[\tau,A_\eta,\varphi]
& = &
\calN_L(\tau)\exp
\Bigg(
-{\tau^2\over 2} \int d^3\tildex \int d^3 \tildey\,
\left[A_\eta(\tildex) G_\eta(\tau,\tildex-\tildey) A_\eta(\tildey)
+
\varphi(\tildex) G_\varphi(\tau,\tildex-\tildey) \varphi(\tildey)
\right.
\non & & \qquad \qquad {}
\left.+
\varphi(\tildex) G_{\varphi\eta}(\tau,\tildex-\tildey) A_\eta(\tildey)
+
A_\eta(\tildex) G_{\eta\varphi}(\tau,\tildex-\tildey)
\varphi(\tildey)\right]
\Bigg)
\ee
Since the normalization factor $\calN_L$ does not depend on the fields,
again we will not consider it.
Consistency under $\tildex\leftrightarrow\tildey$ exchange requires
\be
G_\eta(\tau, \tildex-\tildey)
& = & G_\eta(\tau, \tildey-\tildex)
\\
G_\varphi(\tau, \tildex-\tildey)
& = & G_\varphi(\tau, \tildey-\tildex)
\\
G_{\varphi\eta}(\tau,\tildex-\tildey)
& = &
G_{\eta\varphi}(\tau,\tildey-\tildex)
\ee
Applying the Gauss condition Eq.(\ref{eq:GaussPsi}) on the ansatz,
one obtains the following relationships between the longitudinal kernels
\be
{\partial\over\partial\eta_x}G_\eta(\tau,\tildex-\tildey)
&=& G_{\varphi\eta}(\tau,\tildex-\tildey)
= G_{\eta\varphi}(\tau,\tildey-\tildex)
\label{eq:Gauss1}
\\
{\partial\over\partial\eta_x}G_{\eta\varphi}(\tau,\tildex-\tildey)
&=& G_{\varphi}(\tau, \tildex-\tildey)
\label{eq:Gauss2}
\ee
Comparison between
the left hand side and the right hand side of the functional Schr\"odinger
equation yields the equation for the longitudinal kernel $G_\eta$
\be
-i\partial_\tau (\tau^2 G_\eta(\tau,\tildex-\tildey))
&=&
-{\tau^5} \int d^3\tildeu\, G_\eta(\tau,\tildex-\tildeu)
G_\eta(\tau, \tildeu - \tildey)
\non & & {}
+ \tau^3 \int d^3\tildeu \, d^3\tildev\,
G_{\eta\varphi}(\tau, \tildex-\tildeu) G_\perp(\tildeu-\tildev)
G_{\varphi\eta}(\tau, \tildev-\tildey)
\non & & {}
- {1\over \tau} \nabla_\perp^2 \delta(\tildex-\tildey)
\label{eq:Getaeq_coord}
\ee
where
\be
G_\perp(\tildeu-\tildex)
=
\delta(\eta_u-\eta_x)
{1\over \nabla_\perp^2}
\ee
The equations for $G_\varphi$ and $G_{\eta\varphi}$
are the same as Eq.(\ref{eq:Getaeq}) upon applying the Gauss constraints,
Eqs.(\ref{eq:Gauss1}) and (\ref{eq:Gauss2}).
In the momentum space, the equation for $G_\eta$ becomes
\be
i\partial_\tau\calG_\eta(\tau,\tildek)
&=&
\tau
\left(k_\perp^2 + {k_\eta^2\over \tau^2}\right)
\calG_\eta(\tau,\tildek)
\calG_\eta(\tau, \tildek)
- {1\over \tau}
\label{eq:Getaeq}
\ee
where
\be
\calG_\eta(\tau,\tildek) = {\tau^2\over k_\perp^2} G_{\eta}(\tau,\tildek)
\ee
and we used Eq.(\ref{eq:Gauss1}) to express
$G_{\varphi\eta}$ in terms of $G_{\eta}$.
Let
\be
-i\partial_\tau \psi_\eta(\tau,\tildek) =
\tau
\left(k_\perp^2 + {k_\eta^2\over \tau^2}\right)
\psi_\eta(\tau,\tildek) \calG_\eta(\tau,\tildek)
\label{eq:psieta_def}
\ee
and multiply Eq.(\ref{eq:Getaeq}) with $\psi_\eta$
to get
\be
-i\tau\partial_\tau (\calG_\eta(\tau,\tildek)\psi_\eta(\tau,\tildek))
=
\psi_\eta(\tau,\tildek)
\label{eq:psieta_eq}
\ee
Combining Eq.(\ref{eq:psieta_def}) and Eq.(\ref{eq:psieta_eq}),
one gets
\be
{1\over \tau}\partial_\tau (\tau\partial_\tau \chi(\tau,\tildek))
)
+
\left(k_\perp^2 + {k_\eta^2\over \tau^2}\right)
\chi(\tau,\tildek)
= 0
\ee
where $\chi(\tau,\tildek) = \psi(\tau,\tildek)\calG_\eta(\tau,\tildek)$.
This is again the free field equation for $\chi$.
Stability of vacuum again selects $H^{(1)}_{ik_\eta}(k_\perp\tau)$
\be
\psi_\eta(\tau,\tildek)\calG_\eta(\tau,\tildek) = c'
H^{(1)}_{ik_\eta}(k_\perp\tau)
\ee
where $c'$ is a constant.
Putting this back into Eq.(\ref{eq:psieta_eq}) gives
\be
\psi_\eta(\tau,\tildek)/c'
= -i\tau\partial_\tau H^{(1)}_{ik_\eta}(k_\perp\tau)
= -i\Pi^{(1)}_{ik_\eta}(k_\perp\tau)
\ee
so that
\be
\tau^2 G_{\eta}(\tau,\tildek)
=
{ k_\perp^2 H^{(1)}_{ik_\eta}(k_\perp\tau)
\over
-i\Pi^{(1)}_{ik_\eta}(k_\perp\tau) }
\label{eq:tau2Geta}
\ee
Our choice of the solution $H^{(1)}_{ik_\eta}(k_\perp\tau)$
ensures that the real part of the kernel is positive definite
\be
R_\eta(\tau,\tildek)
& = &
{\rm Re}\left(
\tau^2 G_{\eta}(\tau,\tildek)
\right)
\non
& = &
{ik_\perp^2\over 2}
\left(
{ H^{(1)}_{ik_\eta}(k_\perp\tau)
\over
\Pi^{(1)}_{ik_\eta}(k_\perp\tau) }
-
{ H^{(2)}_{ik_\eta}(k_\perp\tau)
\over
\Pi^{(2)}_{ik_\eta}(k_\perp\tau) }
\right)
\non
& = &
{2k_\perp^2\over \pi}
{ 1 \over \Pi^{(1)}_{ik_\eta}(k_\perp\tau)
\Pi^{(2)}_{ik_\eta}(k_\perp\tau) }
> 0
\ee

The longitudinal part of the vacuum density functional is now
\be
\lefteqn{
\rho_L[\tau_0, A^0_{1\eta},\varphi^0_1, A^0_{2\eta},\varphi^0_2, ]
} &&
\non
& = &
\braket{A^0_{1\eta},\varphi^0_1}{\rm vac, \tau_0}
\braket{\rm vac, \tau_0}{A^0_{2\eta},\varphi^0_2}
\non
& = &
\calN\,
\exp\Bigg(
-
{\tau_0^2\over 2}
\int{d^3\tildek\over (2\pi)^3}\,\left[
\alpha^0_{1\eta}(-\tildek)G_\eta(\tau_0,\tildek)\alpha^0_{1\eta}(\tildek)
+
\alpha^0_{2\eta}(-\tildek)G_\eta^*(\tau_0,\tildek)\alpha^0_{2\eta}(\tildek)\right]
\Bigg)
\label{eq:rhoL}
\non
\ee
where we used the Gauss constraints Eqs.(\ref{eq:Gauss1}) and
(\ref{eq:Gauss2})
to express it solely in terms of the combination
$\alpha^0_{1,2\eta} 
= A^0_{1,2\eta} - \partial_\eta\varphi^0_{1,2}$
or in the momentum space,
\be
\alpha^0_{1,2\eta}(\tildek)
= A^0_{1,2\eta}(\tildek) - ik_\eta\varphi^0_{1,2}(\tildek)
\ee

\section{Wigner Transforms and Propagators}

\subsection{Transverse part}
\label{sec:transverse_props}

With the vacuum wavefunctionals at hand, we can now calculate the
propagators.
The calculation is simpler if one
changes variables to the $r$-$a$ basis (also called the Keldysh rotation)
\be
A_{rT} &=& (A_{1T} + A_{2T})/2
\non
A_{aT} &=& A_{1T} - A_{2T}
\ee
because the separation of the classical and the quantum degrees of freedom
is cleaner in this basis and also because
the ${A_{aT}{-}A_{aT}}$ correlator automatically vanishes.

In this basis, the transverse part of the
generating functional Eq.(\ref{eq:Zfree}) becomes
\be
\lefteqn{Z_{T}[J^T_r, J^T_a]} &&
\non
& = &
\int [dA_{rT}^f][dA_{rT}^0][dA_{aT}^0]\,
\rho_{\rm vac}[\tau_0, A_{rT}^0, A_{aT}^0]\,
\int_{A_{rT}^0}^{A_{rT}^f}\calD A_{rT}
\int_{A_{aT}^0}^{0}\calD A_{rT}
\non & & {} \times
\exp
\left({
i\int_{\tau_0}^\infty d\tau\int d^3\tildex \left(
A_{aT} {\delta\calL_{T}(A_{rT}) \over \delta A_{rT}}
+ J^T_r A_{rT} + J^T_a A_{aT} \right)
-i\int d^3\tildex\, \pi_{rT}^0 A_{aT}^{0}
}\right)
\non
\label{eq:ZJrJa}
\ee
where $A_{rT}^0 = A_{rT}(\tau_0)$, $A_{aT}^{0} = A_{aT}(\tau_0)$ and
\be
\pi_{rT}^0 = \pi_{rT}(\tau_0)
=
\left.
{\partial \calL_{\rm free}(A_{rT}) \over \partial(\partial_\tau A_{rT})}
\right|_{\tau_0}
\ee
is the momentum field at $\tau = \tau_0$.
The term containing $\pi_{rT}^0$
in Eq.(\ref{eq:ZJrJa}) came from integrating by part
in $\tau$. As in Eq.(\ref{eq:Zfree}), we have suppressed the space-time
indices above.

Performing the $\int \calD A_{aT}$ integrals for $\tau > \tau_0$
results in
$ \delta\left[
{\delta\calL_{\rm free}(A_{rT}) \over \delta A_{rT}} + J^T_a
\right] $
which enforces the classical field equation.
Performing
the $\int [dA_{aT}^{0}]$ integral at $\tau_0$ produces
the functional Wigner transform of the vacuum matrix element
\be
\rho_W[\tau, A_{rT}^0, \pi_{rT}^0]
&=&
\int [dA_{aT}^{0}]\,
\rho_{\rm vac}[\tau, A_{rT}^0, A_{aT}^{0}]
\exp\left(
-i\int {d^3\tildek\over (2\pi)^3} \,
A_{aT}^{0}(-\tildek) \pi_{rT}^0(\tildek)
\right)
\ee
It can be shown that the Jacobians coming from integrating over the
$\delta$-functional and changing integration variable from
$A_{rT}^f$ to $\pi_{rT}^0$ (transformation between
the boundary value problem and the initial value problem in classical field
theory)
cancel each other \cite{Jeon:2004dh}.
Hence, the generating functional becomes
\be
\lefteqn{Z_{T}[J_T^r, J_T^a]} &&
\non
& = &
\int [dA_{rT}^0][d\pi_{rT}^0]\,
\rho_W[\tau_0, A_{rT}^0, \pi_{rT}^0]\,
\exp \left({
i\int_{\tau_0}^\infty d\tau\int d^3\tildex
J_T^r A_{rT}[A_{rT}^0, \pi_{rT}^0, J_T^a] }\right)
\label{eq:ZJrJa2}
\ee

The transverse field $A_{rT}[A_{rT}^0, \pi_{rT}^0, J_T^a]$ is the solution
of the equation of motion
\be
\tau \partial^2 A_{rT} + J^a_T = 0
\ee
with the initial conditions $A_{rT}(\tau_0) = A_{rT}^0$,
$\pi_{rT}(\tau_0) = \tau_0\partial_\tau A_{rT}(\tau_0) = \pi_{rT}^0$.
In the momentum space, the full solution is given by
\be
A_{rT}(\tau,\tildek) = A_{rT}^h(\tau,\tildek)
- \int_{\tau_0}^\infty d\tau'\, G_R(\tau,\tau';\tildek)\,
J_T^a(\tau',\tildek)
\ee
where $G_R$ is the retarded Green function defined by
\be
\tau
\left(
{d^2\over d\tau^2}
+
{1\over \tau} {d\over d\tau}
+ k_\perp^2 + {k_\eta^2\over \tau^2}
\right) G_R(\tau,\tau';\tildek) = \delta(\tau - \tau')
\label{eq:GR_def}
\ee
The solution is
\be
G_R(\tau,\tau',\tildek)
= {\pi\over 4i}
\theta(\tau-\tau')
\left(
H_{ik_\eta}^{(1)}(k_\perp\tau)
H_{ik_\eta}^{(2)}(k_\perp\tau')
-
H_{ik_\eta}^{(2)}(k_\perp\tau)
H_{ik_\eta}^{(1)}(k_\perp\tau')
\right)
\label{eq:GRJ_0}
\ee
which can be easily verified by the direct substitution into the defining
equation Eq.(\ref{eq:GR_def}) and using Eq.(\ref{eq:HWronskian}).
Performing Fourier transformation in $k_\eta$, one obtains
(see Refs.\cite{diSessa:1974ve,Svaiter:1989pq} and also Appendix
\ref{app:Mprops})
\be
G_R(\tau,\tau',\eta-\eta',\bfkperp)
= {1\over
2}\theta(\tau-\tau')\theta(\sigma^2)J_0(k_\perp\sqrt{\sigma^2})\label{eq:GrMink}
\ee
with $\sigma^2 = (t-t')^2 - (z-z')^2 = \tau^2 + \tau'^2 -
2\tau\tau'\cosh(\eta-\eta')> 0$.
This coincides with the Minkowski space retarded propagator.
The $r$-$a$ propagator can then be read off from the
$J_T^r{-}J_T^a$ part of the exponent in Eq.(\ref{eq:ZJrJa2}) as
\be
\ave{A_{rT}(\tau,\tildek)A_{aT}(\tau',\tildex=0)}
=
iG_R(\tau,\tau',\tildek)
\ee

To obtain the symmetric propagator
$\ave{A_{rT}(\tau,\tildek)A_{rT}(\tau',-\tildek)}$,
one needs to explicitly evaluate Eq.(\ref{eq:ZJrJa2}).
To do so, we need the homogeneous solution $A_{rT}^h(\tau,\tildek)$
and the Wigner transform $\rho_W[\tau_0,A_{rT}^0,\pi_{rT}^0]$.
With the retarded Green function given in Eq.(\ref{eq:GRJ_0}),
the source term vanishes in the $\tau\to \tau_0$ limit
for both $A_{rT}$ and $\partial_\tau A_{rT}$.
The homogeneous solution satisfying the right boundary conditions is
\be
A_{rT}^h(\tau,\tildek)
& = &
{i\pi\over 4}
\left(
 -H^{(2)}_{ik_\eta}(k_\perp\tau_0) H^{(1)}_{ik_\eta}(k_\perp\tau)
 +H^{(1)}_{ik_\eta}(k_\perp\tau_0) H^{(2)}_{ik_\eta}(k_\perp\tau)
 \right)\pi_{rT}^0(\tildek)
\non & & {}
+
{i\pi\over 4}
\left(
 \Pi^{(2)}_{ik_\eta}(k_\perp\tau_0) H^{(1)}_{ik_\eta}(k_\perp\tau)
 -\Pi^{(1)}_{ik_\eta}(k_\perp\tau_0) H^{(2)}_{ik_\eta}(k_\perp\tau)
\right)A_{rT}^0(\tildek)
\label{eq:classical_AT}
\ee
Using the Wronskian Eq.(\ref{eq:HWronskian}),
it is easy to check that $A_{rT}(\tau,\tildek)$ given in
Eq.(\ref{eq:classical_AT})
does satisfy the initial conditions.
Since the field in the configuration space must be real,
$\pi_{rT}^0(\tildek) = (\pi_{rT}^0(-\tildek))^*$ and $A_{rT}^0(\tildek) =
(A_{rT}^0(-\tildek))^*$.
It is trivial to check that the above solution also satisfies
$A_{rT}^h(\tau,-\tildek) = (A_{rT}^h(\tau,\tildek))^*$
using the Hankel function property
$H^{(1,2)}_{-ik_\eta}(\chi) = (H^{(2,1)}_{ik_\eta}(\chi))^*$.

In the $r$-$a$ basis,
the vacuum density functional
$\rho_{\rm vac}[\tau_0,A^0_{1T},A^0_{2T}]$ in Eq.(\ref{eq:rhoT}) becomes
\be
\rho_{\rm vac}[\tau_0, A_{rT}^0, A_{aT}^{0}]
& = &
\exp\Bigg(
-
\int {d^3\tildek \over (2\pi)^3}\,
R_T(\tau_0,\tildek)\left[A_{rT}^0(-\tildek)A_{rT}^0(\tildek)
+
\frac14 A_{aT}^{0}(-\tildek)A_{aT}^{0}(\tildek)\right]
\non & & \qquad{}
-{i\over 2}
\int {d^3\tildek \over (2\pi)^3}\,
I_T(\tau_0,\tildek)
\left[
A_{aT}^{0}(-\tildek)A_{rT}^0(\tildek)+A_{rT}^0(-\tildek)A_{aT}^{0}(\tildek)
\right]
\Bigg)
\ee

where
$ R_T(\tau_0, \tildek) = {\rm Re}(\tau_0^2 G_T(\tau_0,\tildek)) $
and
$ I_T(\tau_0, \tildek) = {\rm Im}(\tau_0^2 G_T(\tau_0,\tildek)) $
with
$\tau_0^2G_T(\tau_0,\tildek) =
-i\Pi^{(1)}_{ik_\eta}(k_\perp\tau_0)/H^{(1)}_{ik_\eta}(k_\perp\tau_0)$
as given in Eq.(\ref{eq:tau2GT}).
The functional Wigner transform can be carried out by completing the square
in the exponent
\be
\rho_W^T[\tau_0, A^0_{rT}, \pi^0_{rT}]
&=&
\int [dA^{0}_{aT}]\,
\rho_{\rm vac}[\tau_0, A^0_{rT}, A^{0}_{aT}]
\exp\left(
-i\int {d^3\tildek\over (2\pi)^3} \,
A_{aT}^{0}(-\tildek) \pi_{rT}^0(\tildek)
\right)
\non
& = &
\calN\,
\exp
\Bigg(
-
\int {d^3\tildek \over (2\pi)^3}\,\left[
R_T(\tau_0,\tildek)
\left| A_{rT}^0(\tildek) \right|^2
+
{ \left|\pi_{rT}^0(\tildek) + I_T(\tau_0,\tildek)A_{rT}^0(\tildek) \right|^2
\over R_T(\tau_0,\tildek) }
\right]
\Bigg)
\non
\label{eq:rhoWT}
\ee
where $\calN$ is the normalization constant.
This can be interpreted as the probability distribution of the vacuum
fluctuations of the fields $A_{rT}^0$ and $\pi_{rT}^0$ at $\tau_0$.

The generating functional for the symmetric propagator
can be evaluated by completing another square
\be
Z_S[J_T^r]
& = &
\int [dA_{rT}^0][d\pi_{rT}^0]\,
\rho_W[\tau_0, A_{rT}^0, \pi_{rT}^0]\,
\exp \left({
i\int_{\tau_0}^\infty d\tau\int d^3\tildex
J_T^r A_{rT}^h[A_{rT}^0, \pi_{rT}^0] }\right)
\non
& = &
\calN'
\exp\left(
-
{1\over 2}
\int d\tau d\tau'\,
\int {d^3\tildek\over (2\pi)^3}
J_T^r(\tau,-\tildek) J_T^r(\tau',\tildek)
G_S(\tau,\tau';\tildek)
\right)
\ee
where $\calN'$ is another normalization constant.
The symmetric propagator is
\be
\ave{A_{rT}(\tau,\tildek)A_{rT}(\tau',\tildex=0)}
& = &
G_S(\tau,\tau';\tildek)
\non
& = &
\left(\pi\over 8\right)
\left(
H^{(1)}_{ik_\eta}(m_T\tau) H^{(2)}_{ik\eta}(m_T\tau')
+
H^{(2)}_{ik_\eta}(m_T\tau) H^{(1)}_{ik\eta}(m_T\tau')
\right)
\label{eq:TransSym}
\ee
Remarkably, the dependence on $\tau_0$ has gone away.
Similar result was obtained in \cite{Makhlin:2000nx}.
Fourier transforming to $(\tau,\eta,\bfkperp)$ space
yields (see Appendix \ref{app:Hankel} for Hankel function
properties)
\be
G_S(\tau,\tau',\eta-\eta',\bfkperp)
& = &
-\theta(\sigma^2) {1\over 4} Y_0(m_T\sqrt{\sigma^2})
+
\theta(-\sigma^2)
{1\over 2\pi}K_0(m_T\sqrt{-\sigma^2})\label{eq:GsMink}
\ee
This is the same as the Minkowski space symmetric propagator,
further justifying our choice of the solution for
$\psi_T(\tau,\tildek) = H_{ik_\eta}^{(1)}(k_\perp\tau)$
in Section \ref{sec:TransverseVac}.
The same vacuum spectrum and propagators apply to the single component scalar field theory.

\subsection{Longitudinal part}
\label{sec:longi_props}

After the Keldysh rotation
\be
\alpha_{r\eta} &=& (\alpha_{1\eta} + \alpha_{2\eta})/2
\non
\alpha_{a\eta} &=& \alpha_{1\eta} - \alpha_{2\eta}
\ee
the Longitudinal part of the vacuum density functional
Eq.(\ref{eq:rhoL}) becomes
\be
\rho_L[\tau_0,\alpha^0_{r\eta},\alpha^0_{a\eta}]
& = &
\calN\,
\exp\Bigg(
-
\int {d^3\tildek \over (2\pi)^3}\,
R_\eta(\tau_0,\tildek)
\left[
\alpha^0_{r\eta}(-\tildek)\alpha^0_{r\eta}(\tildek)
+
\frac14\alpha^0_{a\eta}(-\tildek)\alpha^0_{a\eta}(\tildek)
\right]
\non & & \qquad{}
-{i\over 2}
\int {d^3\tildek \over (2\pi)^3}\,
I_\eta(\tau_0,\tildek)\left[
\alpha^0_{a\eta}(-\tildek)\alpha^0_{r\eta}(\tildek)
+
\alpha^0_{r\eta}(-\tildek)\alpha^0_{a\eta}(\tildek)\right]\Bigg)
\ee

where
$ R_\eta(\tau_0, \tildek) = {\rm Re}(\tau_0^2 G_\eta(\tau_0,\tildek)) $
and
$ I_\eta(\tau_0, \tildek) = {\rm Im}(\tau_0^2 G_\eta(\tau_0,\tildek)) $
with $\tau_0^2G_\eta$ as given in Eq.(\ref{eq:tau2Geta}).

To perform the functional Wigner transform,
we first add the boundary terms to the exponent
\be
-i\int {d^3\tildek\over (2\pi)^3}\,
\left(
A^0_{a\eta}(-\tildek)\pi^0_{r\eta}(\tildek)
+
\varphi^0_{a\eta}(-\tildek)\pi^0_{r\varphi}(\tildek)
\right)
=
-i\int {d^3\tildek\over (2\pi)^3}\,
\alpha^0_{a\eta}(-\tildek)\pi^0_{r\eta}(\tildek)
\ee
where we used the Gauss law constraint
$\partial_\eta\pi^0_{r\eta} = \pi^0_{r\varphi}$
and integrated by part.
Completing the square then yields the Wigner transform
just as in Eq.(\ref{eq:rhoWT})
\be
\rho_W^L[\tau_0, \alpha^0_{r\eta}, \pi^0_{r\eta}]
& = &
\calN'\,
\exp
\Bigg(
-
\int {d^3\tildek \over (2\pi)^3}\,\left[
R_\eta(\tau_0,\tildek)
\left| \alpha^0_{r\eta}(\tildek) \right|^2
+
\frac{\left|\pi^0_{r\eta}(\tildek) +
I_\eta(\tau_0,\tildek)\alpha^0_{r\eta}(\tildek) \right|^2}
 {R_\eta(\tau_0,\tildek)}\right]
\Bigg)
\non
\label{eq:rhoLW}
\ee
This has the interpretation as the probability distribution of the vacuum
fluctuations of the fields $\alpha^0_{r\eta}$ and $\pi^0_{r\eta}$ at
$\tau_0$.

Since $\rho_W^L$ depends only on the combination
$\alpha^0_{r\eta} = A^0_{r\eta} - ik_\eta\varphi^0_r$,
individual propagators cannot be calculated yet.
Fortunately, there is still one more residual gauge freedom
left in the $A_\tau = 0$ gauge. That is,
the following gauge transformation is still allowed
\be
\bfA'_\perp(\tau,\tildex)
&=& \bfA_\perp(\tau,\tildex) + \nabla_\perp f(\tildex)
\label{eq:gauge_tr_1}
\\
A'_\eta(\tau,\tildex)
&=& A_\eta(\tau,\tildex) + \partial_\eta f(\tildex)
\ee
as long as $f(\tildex)$ is independent of $\tau$.
Since $\nabla_\perp{\cdot}\bfA_\perp = \nabla_\perp^2 \varphi$,
the effect of Eq.(\ref{eq:gauge_tr_1}) on $\varphi$ is
\be
\varphi'(\tau,\tildex)
&=& \varphi(\tau,\tildex) + f(\tildex)
\ee
The momentum field $\pi_\eta$ is unaffected by this transformation.
We can choose $f(\tildex) = -\varphi^0(\tildex)$ so that
\be
\varphi'(\tau_0,\tildex) &=& 0
\\
\inline{and}
A'_\eta(\tau_0,\tildex)
&=& \alpha_\eta(\tildex)
\ee
In this way, $\rho_W^L$ depends solely on $A_{r\eta}^{\prime 0}$ and
$\pi_{r\eta}^0$.

To obtain the longitudinal propagators, we
start with the generating functional containing the vacuum density
functional $\rho_W^L$
\be
Z_L[J_\eta^r, J_\eta^a]
&=&
\int [dA_{r\eta}^0][d\pi_{r\eta}^0]\,
\rho_W^L[\tau_0, A_{r\eta}^0, \pi_{r\eta}^0]\,
\exp\int\left(
i\int J_\eta^r\,A_{r\eta}[A_{r\eta}^0, \pi_{r\eta}^0, J_\eta^a]
\right)
\label{eq:ZL}
\ee
where we now dropped the prime on $A_{r\eta}$. To obtain the propagators,
one needs to find the solutions of the classical field equations
\be
\tau\partial_\tau \left( {1\over \tau}\partial_\tau A_{r\eta} \right)
-
\nabla_\perp^2 A_{r\eta}
+
\partial_\eta \nabla_\perp^2 \varphi_r
+ \tau J_\eta^a
= 0
\label{eq:Areq}
\\
\tau\partial_\tau \left( \tau \partial_\tau \nabla_\perp^2 \varphi_r
\right)
- \partial_\eta^2 \nabla_\perp^2 \varphi_r
+ \nabla_\perp^2 \partial_\eta A_{r\eta}
- \tau J_\varphi^a
= 0
\label{eq:varphireq}
\ee
with the initial conditions determined by $A_{r\eta}^0$ and
$\pi_{r\eta}^0$.
Although $A_{r\eta}$ and $\varphi_r$ themselves cannot be solutions
of the free-field equation, the momentum fields defined by
\be
\pi_{r\eta}  &=& {1\over\tau}\partial_\tau A_{r\eta}
\non
\pi_{r\varphi} &=& -\tau\partial_\tau \nabla_\perp^2\varphi_r
\ee
do satisfy the free-field equation
\be
\tau\left(
\partial_\tau^2 + {1\over \tau}\partial_\tau
- \nabla_\perp^2 - {\partial_\eta^2 \over \tau^2}\right)
\pi_{r\eta,\varphi}
+
\partial_\tau (\tau J_{\eta,\varphi}^a)
= 0
\ee
upon using the Gauss law constraint
$\partial_\eta\pi_{r\eta} = \pi_{r\varphi}$.
The sources must also satisfy
\be\partial_\eta J_\eta^a = J_\varphi^a
\ee
again due to the same Gauss law constraint.

The solution for $\pi_{r\eta}$ is given by
\be
\pi_{r\eta}(\tau,\tildek)
& =&
{i\pi\over 4}
\left(
 \Pi^{(2)}_{ik_\eta}(k_\perp\tau_0) H^{(1)}_{ik_\eta}(k_\perp\tau)
 -\Pi^{(1)}_{ik_\eta}(k_\perp\tau_0) H^{(2)}_{ik_\eta}(k_\perp\tau)
\right)\pi_{r\eta}(\tau_0,\tildek)
\non & & {}
+
{i\pi\over 4}
\left(
 -H^{(2)}_{ik_\eta}(k_\perp\tau_0) H^{(1)}_{ik_\eta}(k_\perp\tau)
 +H^{(1)}_{ik_\eta}(k_\perp\tau_0) H^{(2)}_{ik_\eta}(k_\perp\tau)
\right)(\tau_0\partial_\tau\pi_{r\eta}(\tau_0,\tildek))
\non && {}
-
\int_{\tau_0}^\infty d\tau'\,
G_R(\tau,\tau';\tildek)
\partial_\tau' (\tau' J_\eta^a(\tau',\tildek))
\non
& = &
{i\pi\over 4}
\left(
 \Pi^{(2)}_{ik_\eta}(k_\perp\tau_0) H^{(1)}_{ik_\eta}(k_\perp\tau)
 -\Pi^{(1)}_{ik_\eta}(k_\perp\tau_0) H^{(2)}_{ik_\eta}(k_\perp\tau)
\right)\pi_{r\eta}(\tau_0,\tildek)
\non & & {}
+
{i\pi\over 4}
\left(
 -H^{(2)}_{ik_\eta}(k_\perp\tau_0) H^{(1)}_{ik_\eta}(k_\perp\tau)
 +H^{(1)}_{ik_\eta}(k_\perp\tau_0) H^{(2)}_{ik_\eta}(k_\perp\tau)
\right)(\tau_0\partial_\tau\pi_{r\eta}(\tau_0,\tildek))
\non && {}
+
\int_{\tau_0}^\infty d\tau'\,
(\tau'\partial_{\tau'}G_R(\tau,\tau';\tildek))
J_\eta^a(\tau',\tildek)
+
\tau_0 G_R(\tau,\tau_0;\tildek) J_\eta^a(\tau_0,\tildek)
\label{eq:pieta_sol}
\ee
where the retarded Green function $G_R$ is given in Eq.(\ref{eq:GRJ_0}).
For $\pi_{r\varphi}$, the solution is simply
$\pi_{r\varphi}(\tau,\tildek) = ik_\eta\pi_{r\eta}(\tau,\tildek)$.

The fields are given by
\be
A^r_\eta(\tau,\tildek)
& = &
\int_{\tau_0}^{\tau} d\tau'\, \tau'\, \pi_{r\eta}(\tau',\tildek)
+
A_{r\eta}^0(\tildek)
\label{eq:arsolution}
\ee
and
\be
\varphi_r(\tau,\tildek)
& =&
\int_{\tau_0}^{\tau} {d\tau'\over \tau'}\,
{ik_\eta \pi_{r\eta}(\tau',\tildek)\over k_\perp^2}
\label{eq:varphi_in_pi_eta}
\ee
Putting this back into the equations of motion produces
the following
relationship between $A_{r\eta}^0$ and $\partial_\tau\pi_{r\eta}^0$
\be
A_{r\eta}^0(\tildek) = -{1\over k_\perp^2}
\tau_0 \partial_\tau \pi_{r\eta}(\tau_0,\tildek)
\ee

As before the $r$-$a$ propagator can be read off from the classical
solution
Eq.(\ref{eq:arsolution}) as
\be
\ave{A_{r\eta}(\tau,\tildek)A_{a\eta}(\tau',\tildex=0)}
& = &
{\delta \over i\delta J_a(\tau',\tildek)} A_{r\eta}(\tau,\tildek)
\non
& = &
-i
\theta(\tau-\tau_0)
\int_{\tau_0}^{\tau}d\tau''\tau''\,
\left(\tau' \partial_{\tau'}G_R(\tau'',\tau';\tildek)
+ \tau_0 G_R(\tau'',\tau_0;\tildek)\delta(\tau'-\tau_0)
\right)
\non
\ee
where from Eq.(\ref{eq:GRJ_0})
\be
\tau' \partial_{\tau'}G_R(\tau'',\tau';\tildek)
= {\pi\over 4i}\theta(\tau''-\tau')
\left(
H^{(1)}_{ik_\eta}(k_\perp\tau'')\Pi^{(2)}_{ik_\eta}(k_\perp\tau')
-
H^{(2)}_{ik_\eta}(k_\perp\tau'')\Pi^{(1)}_{ik_\eta}(k_\perp\tau')
\right)
\ee
The $r$-$a$ propagator for the $\varphi$ field can be similarly obtained
using $ik_\eta J_\eta = J_\varphi$
\be
\ave{\varphi_r(\tau,\tildek)\varphi_a(\tau',\tildex=0)}
& = &
-{i\over k_\perp^2}
\theta(\tau-\tau_0)
\int_{\tau_0}^{\tau}{d\tau''\over \tau''}\,
\left(\tau' \partial_{\tau'}G_R(\tau'',\tau';\tildek)
+ \tau_0 G_R(\tau'',\tau_0;\tildek)\delta(\tau'-\tau_0)
\right)
\non
\ee

To obtain the symmetric correlator, we need the homogeneous part of
$A_{r\eta}$ from Eq.(\ref{eq:arsolution})
\be
A_{r\eta}^h(\tau,\tildek)
& = &
\left. A_{r\eta}(\tau,\tildek) \right|_{J_\eta^a=0}
\non
& = &
K(\tau,\tildek) A_{r\eta}^0(\tildek)
+
L(\tau,\tildek) \pi_{r\eta}^0(\tildek)
\ee
where
\be
K(\tau,\tildek)
& =&
1+
{i\pi k_\perp^2\over 4}
\left(
 H^{(2)}_{ik_\eta}(k_\perp\tau_0) R_1^{(1)}(\tau,\tildek)
 -
 H^{(1)}_{ik_\eta}(k_\perp\tau_0) R_1^{(2)}(\tau,\tildek)
\right)
\ee
and
\be
L(\tau,\tildek)
& = &
{i\pi\over 4}
\left(
 \Pi^{(2)}_{ik_\eta}(k_\perp\tau_0) R_1^{(1)}(\tau,\tildek)
 -\Pi^{(1)}_{ik_\eta}(k_\perp\tau_0) R_1^{(2)}(\tau, \tildek)
\right)
\ee
with
\be
R_n^{(1,2)}(\tau,\tildek)
& = &
\int_{\tau_0}^{\tau}d\tau' \tau'^n\, H^{(1,2)}_{ik_\eta}(k_\perp\tau')
\ee
Putting this into Eq.(\ref{eq:ZL}) and completing the square, one gets
\be
\lefteqn{
\ave{A_{r\eta}(\tau,\tildek)A_{r\eta}(\tau',\tildex=0)}
} &&
\non
& = &
{\pi \Pi^{(1)}_{ik_\eta}(k_\perp\tau_0) \Pi^{(2)}_{ik_\eta}(k_\perp\tau_0)
\over 4 k_\perp^2}
+
{\pi k_\perp^2\over 8}
\left(
R_1^{(1)}(\tau,\tildek)R_1^{(2)}(\tau',\tildek)
+
R_1^{(2)}(\tau,\tildek)R_1^{(1)}(\tau',\tildek)
\right)
\non & & {}
-{\pi\over 8}
\left(
\Pi^{(2)}_{ik_\eta}(\tau_0)\left[R_1^{(1)}(\tau,\tildek)
+
R_1^{(1)}(\tau',\tildek)\right]
+
\Pi^{(1)}_{ik_\eta}(\tau_0)\left[R_1^{(2)}(\tau,\tildek)
+
R_1^{(2)}(\tau',\tildek)\right]
\right)
\label{eq:AetaSym}
\ee
The $\pi_{r\eta}{-}\pi_{r\eta}$ correlator can be obtained by applying
$(1/\tau\tau')\partial_\tau\partial_{\tau'}$ to
$\ave{A_{r\eta}(\tau,\tildek)A_{r\eta}(\tau',-\tildek)}$:
\be
\ave{\pi_{r\eta}(\tau,\tildek)\pi_{r\eta}(\tau',\tildex=0)}
& = &
{\pi k_\perp^2\over 8}
\left(
H_{ik_\eta}^{(1)}(k_\perp\tau)H^{(2)}_{ik_\eta}(k_\perp\tau')
+
H_{ik_\eta}^{(2)}(k_\perp\tau)H^{(1)}_{ik_\eta}(k_\perp\tau')
\right)
\label{eq:PiSym}
\ee
The $\varphi_r{-}\varphi_r$ correlator
can be obtained using Eq.(\ref{eq:varphi_in_pi_eta})
\be
\ave{\varphi_r(\tau,\tildex)\varphi_r(\tau',\tildex=0)}
& = &
{\pi k_\eta^2 \over 8 k_\perp^2}
\left(
R_{-1}^{(1)}(\tau,\tildek)R_{-1}^{(2)}(\tau',\tildek)
+
R_{-1}^{(2)}(\tau,\tildek)R_{-1}^{(1)}(\tau',\tildek)
\right)
\label{eq:VarphiSym}
\ee

\section{Comparison with the symplectic product method}
\label{sec:symplectic}

Another method to obtain the propagators amounts to solve the linearized
equation of motion in order to obtain the vacuum spectrum of fluctuations,
the solutions then being normalized with the help of the natural symplectic
product \cite{Dusling:2011rz,Epelbaum:2013waa}.
To see if the symplectic product method is compatible with our approach,
we re-arrange Eq.(\ref{eq:classical_AT}) as
\be
A_{rT}^h(\tau,\tildek)
& = &
{i\pi\over 4}
\left(
 -H^{(2)}_{ik_\eta}(k_\perp\tau_0) \pi_{rT}^0(\tildek)
 + \Pi^{(2)}_{ik_\eta}(k_\perp\tau_0) A_{rT}^0(\tildek)
 \right)
 H^{(1)}_{ik_\eta}(k_\perp\tau)
\non & & {}
+
{i\pi\over 4}
\left(
 H^{(1)}_{ik_\eta}(k_\perp\tau_0) \pi_{rT}^0(\tildek)
 -\Pi^{(1)}_{ik_\eta}(k_\perp\tau_0) A_{rT}^0(\tildek)
\right)
 H^{(2)}_{ik_\eta}(k_\perp\tau)
\label{eq:classical_AT2}
\ee

The vacuum functional $\rho_W^T$ in Eq.(\ref{eq:rhoWT})
has the exponent
\be
S_T
& = &
-\int {d^3\tildek\over (2\pi)^3}
\left( R_T|A_{rT}^0|^2 + |\pi_T^0 + I_T A_{rT}^0|^2/R_T \right)
\label{eq:ST1}
\ee
where $R_T$ and $I_T$ are the real and the imaginary part
of the kernel
$\tau_0^2 G_T(\tau_0,\tildek)$ as given in (\ref{eq:tau2GT}).
Since $A_{rT}^0(\tildex)$ and $\pi_{rT}^0(\tildex)$ are real,
we have $A_{rT}^0(-\tildek) = (A_{rT}^0(\tildek))^*$
and
$\pi_{rT}^0(-\tildek) = (\pi_{rT}^0(\tildek))^*$.
Hence the real part of $A_{rT}^0(\tildek)$ and $\pi^0_{rT}(\tildek)$
must be an even function of $\tildek$
and the imaginary part must be an odd function of $\tildek$.
The real and imaginary parts of $\tau_0^2 G_T(\tau_0,\tildek)$ are
also even functions of $\tildek$ because
$H_{-ik_\eta}^{(1)}(k_\perp\tau)
= e^{-\pi k_\eta} H_{ik\eta}^{(1)}(k_\perp\tau)$
(see appendix section \ref{app:Hankel}).
Hence
under the $\int d^3\tildek$ integral, we can replace
\be
R_T |A_{rT}^0|^2 + \frac{|\pi^0_{rT} + I_TA_{rT}^0|^2}{R_T}
\to
\frac{|\pi_{rT}^0 -i (R_T + iI_T)A_{rT}^0|^2}{R_T}
\ee
because the difference is an odd function of $\tildek$.
This allows us to re-express the exponent in $\rho_T^W$ as
\be
S_T
& = &
-2
\int {d^3\tildek\over (2\pi)^3}
\left|
\pi_{rT}^0(\tildek)
a^*_{\tildek}(\tau_0)
-
A_{rT}^0(\tildek) e^*_{\tildek}(\tau_0)
\right|^2
\non
\ee
where used the fact that
$1/R_T = \pi
H_{ik_\eta}^{(1)}(k_\perp\tau_0)H_{ik_\eta}^{(2)}(k_\perp\tau_0)/2$
and defined
\be
a_{\tildek}(\tau)
& = &
\sqrt{\pi\over 4}
e^{\pi k_\eta/2} H_{ik_\eta}^{(2)}(k_\perp\tau)
\label{eq:atildektau}
\ee
and
\be
e_{\tildek}(\tau) = \tau\partial_\tau a_{\tildek}(\tau)
\label{eq:etildektau}
\ee
For large $\tau$, $a_{\tildek}(\tau)$
\be
a_{\tildek}(\tau) \sim {1\over \sqrt{2 k_\perp\tau}}
e^{-i(k_\perp\tau - \pi/4)}
\ee
Hence $a_{\tildek}(\tau)$ can be regarded as the positive
frequency solution.
Since
\be
a^*_{-\tildek}(\tau)
=
\sqrt{\pi\over 4}
e^{-\pi k_\eta/2} H_{ik_\eta}^{(1)}(k_\perp\tau)
\ee
the Wronskian Eq.(\ref{eq:HWronskian}) is equivalent to
the symplectic product
\be
e^*_{-\tildek}(\tau)a_{\tildek}(\tau)
-
e_{\tildek}(\tau)a^*_{-\tildek}(\tau)
= i
\ee
In terms of $a_{\tildek}(\tau)$, the homogeneous solution is
\be
A_{rT}^h(\tau,\tildek)
& = &
c_\tildek a_{\tildek}(\tau)
+
c^*_{-\tildek} a^*_{-\tildek}(\tau)
\ee
where
\be
c_\tildek
& = &
i
\left(
\pi_{rT}^0(\tildek)
a^*_{ik_\eta}(k_\perp\tau_0)
-
A_{rT}^0(\tildek)
e^*_{ik_\eta}(k_\perp\tau_0)
\right)
\label{eq:ctildek}
\ee
In terms of $c_\tildek$, the exponent of $\rho_T^W$ becomes
much simpler
\be
S_T =
-2\int {d^3\tildek\over (2\pi)^3}\, c_\tildek c^*_\tildek
\label{eq:ST2}
\ee
which implies
\be
\ave{c_{\tildek}c^*_{\tildek'}} = {1\over
2}(2\pi)^3\delta(\tildek-\tildek')
\ee
This is identical to the correlation function obtained with the symplectic
product approach \cite{Dusling:2011rz,Epelbaum:2013waa}.

For the longitudinal part, the exponent in $\rho_L^W$ is
\be
S_L =
- \int {d^3\tildek\over (2\pi)^3}\,
\left(R_\eta |A_{r\eta}^0|^2
+
{ |\pi_{r\eta}^0 + I_\eta A_{r\eta}^0|^2\over R_\eta }
\right)
\ee
where $R_\eta$ and $I_\eta$ is the real and the imaginary part of
${ik_\perp^2 H^{(1)}_{ik_\eta}(k_\perp\tau_0) /
\Pi_{ik_\eta}^{(1)}(k_\perp\tau)}$.
This can be re-expressed as
\be
S_L
& = &
- \int {d^3\tildek\over (2\pi)^3}\,
{|\pi_{r\eta}^0 -i (R_\eta + iI_\eta) A_{r\eta}^0|^2\over R_\eta }
\ee
again using the fact that the real parts of $A_{r\eta}^0$ and
$\pi_{r\eta}^0$ are even
functions of $\tildek$ and the imaginary parts are odd functions of
$\tildek$.
In terms of $a_{\tildek}(\tau)$ and $e_{\tildek}(\tau)$ defined in
Eqs.(\ref{eq:atildektau}) and (\ref{eq:etildektau}),
we have
\be
R_\eta + iI_\eta = ik_\perp^2
{a^*_{-\tildek}(\tau_0)\over e^*_{-\tildek}(\tau_0)}
\ee
and
\be
R_\eta = {k_\perp^2 \over 2
e_{\tildek}(\tau_0) e^*_{-\tildek}(\tau_0)}
\ee
Hence
\be
S_L
& = &
-
\int {d^3\tildek\over (2\pi)^3}\,
{|\pi_{r\eta}^0 -i (R_\eta + iI_\eta) A_{r\eta}^0|^2\over R_\eta }
\non
& = &
-2\int {d^3\tildek\over (2\pi)^3}\,
{1\over k_\perp^2}
\left|e^*_{-\tildek}\pi_{r\eta}^0(\tildek)
+ k_\perp^2 a^*_{-\tildek} A_{r\eta}^0(\tildek)\right|^2
\ee
The same combination in the integrand occurs in
(c.f.~Eq.(\ref{eq:pieta_sol}))
\be
\pi_{r\eta}(\tau,\tildek)
=
d_{\tildek} a_{\tildek}(\tau)
+ d^*_{-\tildek} a_{-\tildek}^*(\tau)
\ee
where
\be
d_{\tildek} =
e^*_{-\tildek}(\tau_0)\pi_{r\eta}^0(\tildek)
+
k_\perp^2 a^*_{\tildek}(\tau_0)  A_{r\eta}^0(\tildek)
\ee
Again, the exponent simplifies in terms of $d_{\tildek}$
\be
S_L =
-2\int {d^3\tildek\over (2\pi)^3}\,
{1\over k_\perp^2}\, d_{\tildek}d^*_{\tildek}
\ee
which results in
\be
\ave{d_{\tildek}d^*_{\tildek'}}
=
{k_\perp^2 \over 2} (2\pi)^3\delta(\tildek-\tildek')
\label{eq:ddstar}
\ee
which is again equivalent to the correlation function used
in the symplectic method.

Two remarks are in order here.
The first one concerns the number of independent stochastic degrees
of freedom.
If one compares Eq.(\ref{eq:ST1}) and Eq.(\ref{eq:ST2}),
it looks like the number of stochastic degrees of freedom
is reduced by a factor of 2 between Eq.(\ref{eq:ST1})
and Eq.(\ref{eq:ST2}). This is not so. Even though it looks like there
are 4 functional degrees of freedom in Eq.(\ref{eq:ST1}),
they are not complete. Recall that the real part
of $A_{rT}^0$ and $\pi_{rT}^0$ is an even function of $\tildek$
and the imaginary part is an odd function. Thus, each cover only
1/2 of the function space.
On the other hand, there is no such restriction on
$c_\tildek$ in Eq.(\ref{eq:ctildek}) and hence there is no contradiction.
The second remark is that for numerical implementation
Eq.(\ref{eq:ST2}) is much more suited than Eq.(\ref{eq:ST1}), which is what
has been used in \cite{Dusling:2010rm}.

\section{Conclusion}

In this paper, we have derived in Milne space the initial density matrix of
an abelian gauge field theory
in vacuum, by solving the corresponding Schr\"odinger equation. This
allowed us
to compute the Wigner transform of the density matrix, leading to a simple
expression for the transverse
and longitudinal parts of the generating functional of the Closed Time Path
formalism. From
there can finally be deduced the propagators of the theory. The key
finding of
this study is the perfect agreement between the results obtained with the
functional
Schr\"odinger approach and those derived with the symplectic product based
method \cite{Epelbaum:2013waa} -- as demonstrated in section
\ref{sec:symplectic}.
An important though challenging extension of this work would be to compute
the initial density matrix of the Wave-functional corresponding
to the so-called Glasma state \cite{McLerran:1993ni, McLerran:1993ka,
McLerran:1994vd},
and to compare with the result of \cite{Epelbaum:2013waa} in this
interacting case. 

Establishing with the Schr\"odinger functional approach the Glasma initial
condition is of crucial importance for dealing with the problem of
thermalization in heavy ion collision. Indeed, the quark-gluon plasma
created in these collisions seem to behave like a nearly perfect fluid
\cite{Luzum:2008cw,Romatschke:2009im} on very short time scales, as
corroborated by numerous hydrodynamical studies that are able to reproduce
the experimental data collected at the RHIC and LHC colliders. But a first
principle proof of the applicability of hydrodynamics for the quark-gluon
plasma -- that is initially not thermal\cite{Krasnitz:1998ns,Lappi:2006fp}
-- has so far been lacking. An important building-block towards the
completion of this proof is the interacting counterpart of the present
work, derived with the symplectic product based approach
\cite{Epelbaum:2013waa}, that has recently been used numerically to obtain
promising results \cite{Gelis:2013rba}. Improving our theoretical
understanding of the Glasma initial condition with the Schr\"odinger
functional approach would therefore be of great help in the path towards
the full understanding of the nearly perfect fluidity of the quark-gluon
plasma.

Another challenging extension is the NLO contribution to the vacuum wavefunctionals.
At the leading order, there is no real difference between 
the Abelian vacuum and the non-Abelian vacuum. However, the NLO contribution must
be different. In view of recent developments on the NLO JIMWLK equations and factorizations
\cite{Balitsky:2013fea,Kovner:2013ona,Kovner:2014lca,Caron-Huot:2013fea,Caron-Huot:2015bja},
this would be an important step to establish firm connection between
the current Schwinger-Keldysh approach and the CGC approaches.

\section*{Acknowledgements}

S.J. and T.E.~thanks A.~Maloney, F.~Gelis, K.~Dusling, R.~Venugopalan, G.~Moore and C.~Gale
for discussions and suggestions over the course of developing
the current manuscript.
S.J.~and T.E.~are supported in part by the Natural Sciences and
Engineering Research Council of Canada and by le Fonds
Nature et Technologies of Qu\'ebec.

\appendix

\section{Hankel function properties}
\label{app:Hankel}

The definition of the Hankel functions used here are
\be
H^{(1)}_{ik_\eta}(\chi)
=
-{ie^{\pi k_\eta/2} \over \pi}
\int_{-\infty}^{\infty} d\zeta\, e^{i\chi\cosh\zeta \mp ik_\eta \zeta}
\ee
and
\be
H^{(2)}_{ik_\eta}(\chi)
&=&
{ie^{-\pi k_\eta/2} \over \pi}
\int_{-\infty}^{\infty} d\zeta\, e^{-i\chi\cosh\zeta \pm ik_\eta \zeta}
\ee
with real $\chi$. Both signs of $ik_\eta\zeta$ leads to the same function
because they are related by the simple
change of the integration variable from $\zeta$ to
$-\zeta$.
{}From these definitions, it is easy to see
\be
&
\left( H^{(1)}_{ik_\eta}(\chi) \right)^*
=
e^{\pi k_\eta} H^{(2)}_{ik_\eta}(\chi)
\\
&
H^{(1)}_{-ik_\eta}(\chi) = e^{-\pi k_\eta} H^{(1)}_{ik_\eta}(\chi)
= (H^{(2)}_{ik_\eta}(\chi))^*
\ee
Using the definitions, one can perform
Fourier transformation of the product of Hankel functions.
Let
\be
I_H(\tau,\tau',\eta-\eta')
& = &
\int_{-\infty}^{\infty} {dk_\eta\over 2\pi}\,
e^{ik_\eta (\eta - \eta')}
H^{(1)}_{ik_\eta}(m\tau)
H^{(2)}_{ik_\eta}(m\tau')
\non
& = &
{1\over \pi^2}
\int_{-\infty}^\infty d\zeta\,
e^{im\tau\cosh(\zeta-\eta)}
e^{-im\tau'\cosh(\zeta-\eta')}
\non
& = &
{1\over \pi^2}
\int_{-\infty}^\infty d\zeta\,
e^{im( (t-t')\cosh\zeta-(z-z')\sinh\zeta )}
\ee
where $t = \tau\cosh\eta$ and $z = \tau\sinh\eta$.
If $(t-t')^2 > (z-z')^2$, then the exponent can be combined into
\be
(t-t')\cosh\zeta-(z-z')\sinh\zeta
= {\rm sign}(t-t')\sigma \cosh(\zeta-\zeta_0)
\ee
where $ \zeta_0 = \tanh^{-1}((z-z')/(t-t')) $  and
\be
\sigma &=& \sqrt{(t-t')^2 - (z-z')^2}
\non
&=&
\sqrt{ \tau^2 + \tau'^2 - 2 \tau\tau'\cosh(\eta-\eta')}
\ee
Hence
\be
I_H(\tau,\tau',\eta-\eta')
=
\left\{
\begin{array}{ll}
{i\over \pi} H_0^{(1)}(m\sigma) & \hbox{if $t > t'$}
\\
-{i\over \pi} H_0^{(2)}(m\sigma) & \hbox{if $t' > t$}
\end{array}
\right.
\ee
Similarly one can show
\be
I'_H(\tau,\tau';\eta-\eta')
& =&
\int_{-\infty}^{\infty} {dk_\eta\over 2\pi}\,
e^{ik_\eta (\eta - \eta')}
H^{(2)}_{ik_\eta}(m\tau)
H^{(1)}_{ik_\eta}(m\tau')
\non
& = &
(I_H(\tau,\tau';\eta-\eta'))^*
\ee
and since $H_0^{(1)}(x) = J_0(x) + iY_0(x)$, we get
\be
I_H + I'_H = -{2\over \pi}Y_0(m\sigma)
\ee
as well as
\be
I_H - I'_H = {2i\over \pi} J_0(m\sigma)
\ee

\section{Transformation of Minkowski space propagators}
\label{app:Mprops}
We want to show the equivalence between the symmetric and retarded
propagators in Milne space and the usual Minkowskian ones. Starting from
the symmetric propagator in the Minkowskian coordinate system
\be
G_S(k) = \pi\delta(k_0^2 - E_k^2)
\ee
we can re-express it in the following way, depending on the region of the
light cone considered
(1) In the forward light cone region, where $t^2 - z^2 > 0$ and $t>0$, a
straightforward calculation leads to
\be
G_S
& = &
-{1\over 4}Y_0(m_T\tau)
\ee
(2) In the backward light cone region, where $t^2 - z^2 > 0$ and $t<0$, one
easily finds the same result as in the forward light cone region
\be
G_S
& = &
-{1\over 4}Y_0(m_T\tau)
\ee
(3)
In the right side of the light cone, where $z^2 > t^2$ and $z > 0$, one can
use the fact that
\be
k^z z \pm E_k t
& = &
m_T\sigma \sinh(y\pm\eta)
\ee
to show that
\be
G_S
& = &
{1\over 2\pi}K_0(m_T\tau)
\ee
(4)
In the left side of the light cone, where $z^2 > t^2$ and $z < 0$, one also
finds
\be
G_S
& = &
{1\over 2\pi}K_0(m_T\tau)
\ee
which gives back (\ref{eq:GsMink}) as anticipated.

Now for the retarded propagator, we  start with
\be
G_R = \theta(t){\sin(Et)\over E}
\ee
which can be re-expressed into
\be
G_R & = &
\int {dk^z\over 2\pi}
{e^{-ik^z z + iEt} - e^{ik^z z-iEt}\over 2iE}
\ee
Consider the following two cases

(1) If we assume $|z| > |t|$, then
\be
k^z z - Et & = &
{\rm sign}(z)m_T\sqrt{-\sigma^2}
\sinh(y - {\rm sign}(z)\eta)
\ee
leads to
\be
G_R = \int {dy\over 2\pi} \sin(x \sinh(y)) = 0
\ee
(2) If we now assume $|t| > |z|$, then
\be
k^z z - Et & = &
-m_T\sqrt{\sigma^2}{\rm sign}(t)
\cosh(y - \sign(t)\eta)
\ee
leads to
\be
G_R
& = &
{1\over 2} J_0(m_T\sqrt{\sigma^2})
\ee
as announced in (\ref{eq:GrMink}).

\bibliographystyle{aipnum4-1}
\bibliography{bigrefs.bib}
\end{document}

%% file: my.tex
\def\blfootnote{\xdef\@thefnmark{}\@footnotetext}

\long\def\symbolfootnote[#1]#2{\begingroup%
\def\thefootnote{\fnsymbol{footnote}}\footnote[#1]{#2}\endgroup}

\newcommand{\be}{\begin{eqnarray}}
\newcommand{\ee}{\end{eqnarray}}
\newcommand{\ben}{\begin{eqnarray*}}
\newcommand{\een}{\end{eqnarray*}}

\newcommand{\bcent}{\begin{center}}
\newcommand{\ecent}{\end{center}}
\newcommand{\benum}{\begin{enumerate}}
\newcommand{\eenum}{\end{enumerate}}
\newcommand{\bdesc}{\begin{description}}
\newcommand{\edesc}{\end{description}}

\newcommand{\bitem}{\begin{itemize}}
\newcommand{\eitem}{\end{itemize}}

\newcommand{\bitemset}[1]{\begin{itemize}\addtolength{\itemsep}{#1}}
\newcommand{\biteml}{\begin{itemize}\addtolength{\itemsep}{+0.1cm}\large}
\newcommand{\bitemL}{\begin{itemize}\addtolength{\itemsep}{+0.1cm}\Large}
\newcommand{\bitemh}{\begin{itemize}\addtolength{\itemsep}{+0.1cm}\huge}
\newcommand{\bitemH}{\begin{itemize}\addtolength{\itemsep}{+0.1cm}\Huge}
\newcommand{\bitemn}{\begin{itemize}\addtolength{\itemsep}{+0.1cm}\normalsize}

\newcommand{\bquote}{\begin{quote}}
\newcommand{\equote}{\end{quote}}
\newcommand{\bhalfp}{\begin{minipage}{0.45\textwidth}}
\newcommand{\ehalfp}{\end{minipage}}
\newcommand{\bhead}{\begin{center}\bf \Large}
\newcommand{\ehead}{\end{center}\bigskip}
\newcommand{\non}{\nonumber\\}

\newcommand{\inline}[1]{\noalign{\hbox{#1}}}

 \newcommand{\bfe}{{\bf e}}

 \newcommand{\bfk}{{\bf k}}

 \newcommand{\bfx}{{\bf x}}

 \newcommand{\bfA}{{\bf A}}

 \newcommand{\bfkperp}{{\bf k}_\perp}

 \newcommand{\bfxperp}{{\bf x}_\perp}

 \newcommand{\bfAperp}{{\bf A}_\perp}

 \newcommand{\hatbfk}{\hat{\bf k}}

 \newcommand{\hatk}{{\hat{k}}}

 \newcommand{\hatH}{{\hat{H}}}

 \newcommand{\hatO}{{\hat{O}}}

 \newcommand{\hatT}{{\hat{T}}}
 \newcommand{\hatU}{{\hat{U}}}

 \newcommand{\hatrho}{{\hat{\rho}}}

 \newcommand{\tildek}{{\tilde{k}}}

 \newcommand{\tildeu}{{\tilde{u}}}
 \newcommand{\tildev}{{\tilde{v}}}
 
 \newcommand{\tildex}{{\tilde{x}}}
 \newcommand{\tildey}{{\tilde{y}}}
 
 \newcommand{\calD}{{\cal D}}

 \newcommand{\calG}{{\cal G}}
 \newcommand{\calH}{{\cal H}}

 \newcommand{\calL}{{\cal L}}
 \newcommand{\calM}{{\cal M}}
 \newcommand{\calN}{{\cal N}}

 \newcommand{\bra}[1]{\langle {#1} |}
 \newcommand{\ket}[1]{| {#1} \rangle}
 \newcommand{\braket}[2]{
              \left\langle
	      \left. {#1} \vphantom{#1} \vphantom{#2} \right|
	      {#2}\right\rangle}

 \newcommand{\sign}{\hbox{sign}}

 \newcommand{\ave}[1]{{\langle {#1} \rangle}}

\newcommand{\bmat}[1]{\left(\begin{array}{#1}}
\newcommand{\emat}{\end{array}\right)}

\usepackage[usenames,dvipsnames,svgnames,table]{xcolor}
\usepackage{graphicx}
\usepackage{hyperref}
\usepackage{amsmath}
\usepackage{slashed}
\usepackage[normalem]{ulem}

\usepackage[normalem]{ulem} 